\documentclass{aa}
\usepackage{graphicx}
\usepackage{txfonts}
\usepackage[draft]{hyperref}
\usepackage{color}
\usepackage{natbib}
\usepackage{placeins}

\bibpunct{(}{)}{;}{a}{}{,} 

\newcommand{\Msun}{\ensuremath{M_{\odot}}}
\newcommand{\sub}[2]{\ensuremath{#1_{\mathrm{#2}}}}

\newcommand{\unit}[2]{\ensuremath{\textrm{#1}^{#2}}}
\newcommand{\half}{\frac{1}{2}}
\newcommand{\vect}[1]{\mathbf{#1}}

\definecolor{robyncolor}{RGB}{127,0,0}

\begin{document}

\title{Stream-subhalo interactions in the Aquarius simulations}

\author{R.E. Sanderson \inst{1,2,3}\fnmsep\thanks{\email{robyn@caltech.edu}}
\and
C. Vera-Ciro \inst{4}
\and
A. Helmi \inst{5}
\and
J. Heit \inst{5}
}

\institute{
TAPIR, Caltech, MC 350-15, 1200 E California Blvd, Pasadena, CA, 91125, USA
\and
Department of Astronomy, Columbia University, Pupin Physics Laboratory, 550 West 120th Street, New York, NY 10027, USA
\and
NSF Astronomy and Astrophysics Postdoctoral Fellow
\and
Departamento de Ciencias B\'asicas, Universidad de Medell\'in, Cra 87 N 30-65, Medell\'in, Colombia
\and
Kapteyn Astronomical Institute, P.O. Box 800, 9700 AV Groningen, The Netherlands
}

\date{Version \today}


\abstract{}{We perform the first self-consistent measurement of the rate of interactions between stellar tidal streams created by disrupting satellites and dark subhalos in a cosmological simulation of a Milky-Way-mass galaxy.} {Using a retagged version of the Aquarius A dark-matter-only simulation, we selected 18 streams of tagged star particles that appear thin at the present day and followed them from the point their progenitors accrete onto the main halo, recording in each snapshot the characteristics of all dark-matter subhalos passing within several distance thresholds of any tagged star particle in each stream. We considered distance thresholds corresponding to constant impact parameters (1, 2, and 5 kpc), as well as those proportional to the region of influence of each subhalo (one and two times its half-mass radius $r_{1/2}$). We then measured the age and present-day, phase-unwrapped length of each stream in order to compute the interaction rate in different mass bins and for different thresholds, and compared these to analytic predictions from the literature.} {We measure a median rate of $1.5^{+3.0}_{-1.1}\ (9.1^{+17.5}_{-7.1},\ 61.8^{+211}_{-40.6})$ interactions within 1 (2, 5) kpc of the stream per 10 kpc of stream length per 10 Gyr. Resolution effects (both time and particle number) affect these estimated rates by lowering them.}{}

\keywords{Galaxy: halo --
 Galaxy: kinematics and dynamics --
 (cosmology:) dark matter --
 Galaxy: structure --
 galaxies: interactions --
 methods: numerical}

\maketitle


\section{Introduction}
A key prediction of the concordance model of the universe is the
existence of structure in the dark matter distribution at nearly all
scales, limited only by the intrinsic temperature of the candidate
dark matter particle at the smallest scales and causality at the
largest scales \citep{2005PhR...405..279B}. At the galactic scale this
structure manifests as bound subhalos orbiting within the
gravitational potential of their host, with the largest mass
fraction in subhalos generally occuring near the virial radius of the
main halo \citep{Springel2008}. However it is also clear that many of
these subhalos, if they exist, must not host luminous galaxies, since
their number in cold-dark-matter simulations greatly
exceeds the number of satellite galaxies observed around the Milky Way
(MW) and other galaxies in nature, the so-called missing satellite
problem \citep[and Moore 1999]{1999ApJ...522...82K}. Solutions to this
discrepancy fall into two broad categories: those that modify the
properties of dark matter to raise the minimum size of substructures
to match observed satellites, and those in which the dark matter is cold (CDM)
 and baryonic processes select some fraction of subhalos to
host galaxies. Solutions that adjust the baryons rather than the dark
matter thus also predict the existence of many ``dark subhalos''
devoid of baryonic matter, orbiting nearly invisibly in galactic
halos. Determining whether these dark subhalos exist is thus an
important test to discriminate between CDM and other models of dark
matter.

One possible observable signature of dark subhalos is the imprint they
leave on stellar tidal streams, the unbound remnants of accreted
satellite galaxies. These satellite galaxies initially inhabit the
subhalos that managed to retain their gas and form stars before 
accreting onto their host galaxy. Many tidal streams have already
been found \citep{1999HelmiNat, 2001ApJ...547L.133I,
  2001ApJ...554L..33V, 2003AJ....126.2385O, 2003ApJ...588..824Y,
  2004ApJ...605..575Y, 2006ApJ...642L.137B, 2006ApJ...645L..37G,
  2006ApJ...636L..97D, 2006ApJ...651L..33L, 2006ApJ...639L..17G,
  2006ApJ...643L..17G, 2007ApJ...658..337B, 2009ApJ...698..865K,
  2009ApJ...693.1118G, 2012ApJ...760L...6B, 2013ApJ...765L..39M} and
still more are thought to exist
\citep[e.g.][]{1999Helmi,Helmi2011,Gomez2013}. Tidal streams are
kinematically cold compared to the total phase-space distribution of
their host, since they come from less massive progenitors, and
their stars orbit as test particles in the host after being
unbound by tides from their progenitors. A tidal stream will thus
become even colder as it ages, since its stars conserve their total
phase-space volume and slight differences in their orbits cause their
positions to spread out over time \citep[e.g.][]{1999Helmi}. This
elongation of the volume in physical space makes streams sensitive probes of
the lumpiness of the mass distribution, since interactions with dark
subhalos will disturb the long (10s to 100s of kpc), thin (physical
width of 10s to 100s of pc, velocity dispersion of a few km
\unit{s}{-1}) stream of stars
\citep{Johnston2002,Ibata2002,Mayer2002,Penarrubia2006,Carlberg2009,Yoon2011}. The
small velocity spread and width of a stream mean that even small
disturbances can be detected, while its long length gives it a large
cross-section to interactions with the more isotropic population of
dark substructure. Additionally, the arrangement of the stream stars
along neighboring orbits will magnify alterations over time, since a
local encounter with a subhalo will affect a subset of stream stars
with similar orbital phases. Thus for example a ``gap'' in a stream
caused by scattering from a subhalo will widen over time as the
discrepancy between orbits at the edges of the gap increases
\citep{Carlberg2009,Yoon2011}.

Previous work on stream-subhalo interactions has fallen into two categories. \citet{Carlberg2009,2013ApJ...775...90C,2015ApJ...800..133C,2015ApJ...808...15C} and \citet{2015arXiv150705625E} considered the effects of these encounters on idealized streams on relatively circular orbits to determine the characteristics of their signatures and the mass range of subhalos likely to create observable gaps.  Other work by \citet[][hereafter YJH]{Yoon2011}, \citet{Carlberg2012},  \citet{2014ApJ...788..181N}, and \citet{2015ApJ...803...75N} inserted simulated globular-cluster-like streams into dark matter halos based on cosmological simulations to estimate how often such interactions would occur in CDM halos. However, there has been no attempt to date to self-consistently measure the interaction rate between streams from accreted luminous substructures and nonluminous subhalos generated from a single cosmological simulation. That is the goal of this work.

To calculate the predicted frequency of both streams and subhalos, we use one of the Aquarius simulations of a Milky-Way-like dark matter halo \citep{Springel2008}, a portion of the Millenium II dark-matter-only cosmological simulation \citep{2009MNRAS.398.1150B} that has been resimulated at higher resolution. Specifically we use the simulation of Halo Aq-A-2 ($M_{200} = 1.842 \times 10^{12}\ M_{\odot}$) at resolution level 2 ($m_p = 1.37 \times 10^4\ M_{\odot}$). This simulation resolves dark substructures down to about $10^6\ M_{\odot}$ at size scales of about 0.3 kpc. To create stellar streams from some of the substructures, we tag some of the dark matter particles as ``stars" according to the procedure described in Section \ref{sec:retagging}. We then follow one such set of stars, which at the present day in the simulation forms a long thin stream, and track the close encounters between these stars and the many untagged dark subhalos in each snapshot of the simulation, from the infall of the stream's progenitor to the present day (Section \ref{sec:exampleStream}) and examine the rate of interactions and the evolution of the stream in phase space and constants of motion. Finally, we select 18 streams that still appear long and thin at the present day and track their interactions with subhalos in the same way to calculate an average interaction rate for a stream of a given age and present day length (Section \ref{sec:intxnStats}). In Section \ref{sec:discussion} we compare the rate to estimates from the literature and in Section \ref{sec:concl} we summarize our findings.

\begin{figure}
  \centering
\includegraphics[width=0.43\textwidth]{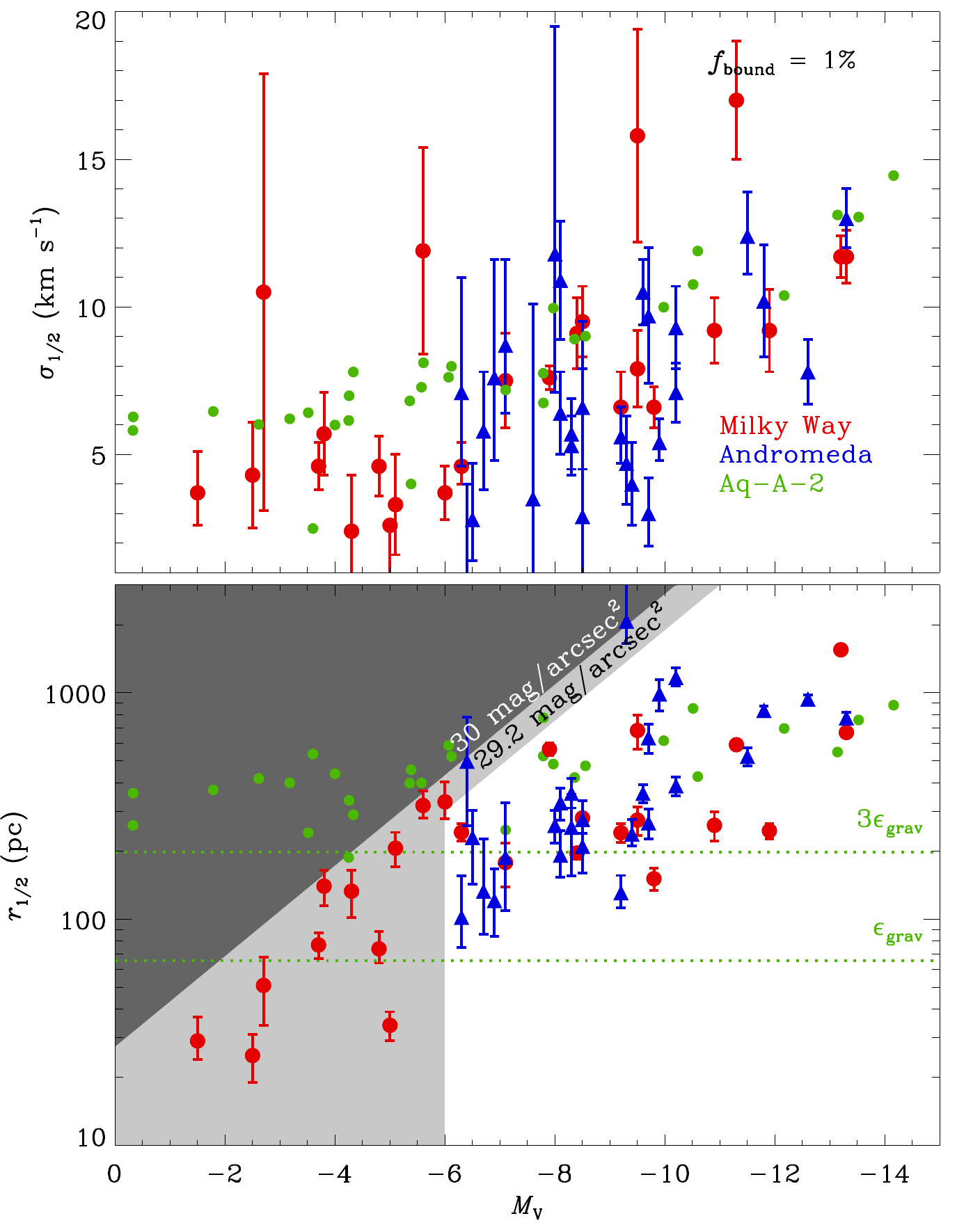}
  \caption{Luminosity-velocity dispersion and luminosity-size relations of Aq-A-2 ``satellites'' still bound at the present day (green points, produced by tagging the 1\% most bound DM particles at infall) compared to known MW (red) and M31 (blue) satellites \citep{2015ApJ...799L..13C}. The green lines show multiples of the softening length used in the dark-matter-only N-body simulation. Compare with Figure 4 of \citet{2010MNRAS.406..744C}. The dark grey and light grey regions are surface brightness limits from SDSS and PandAs, respectively.}
  \label{fig:lfunc}
\end{figure}

 \begin{figure*}
   \centering
 \includegraphics[width=0.95\textwidth]{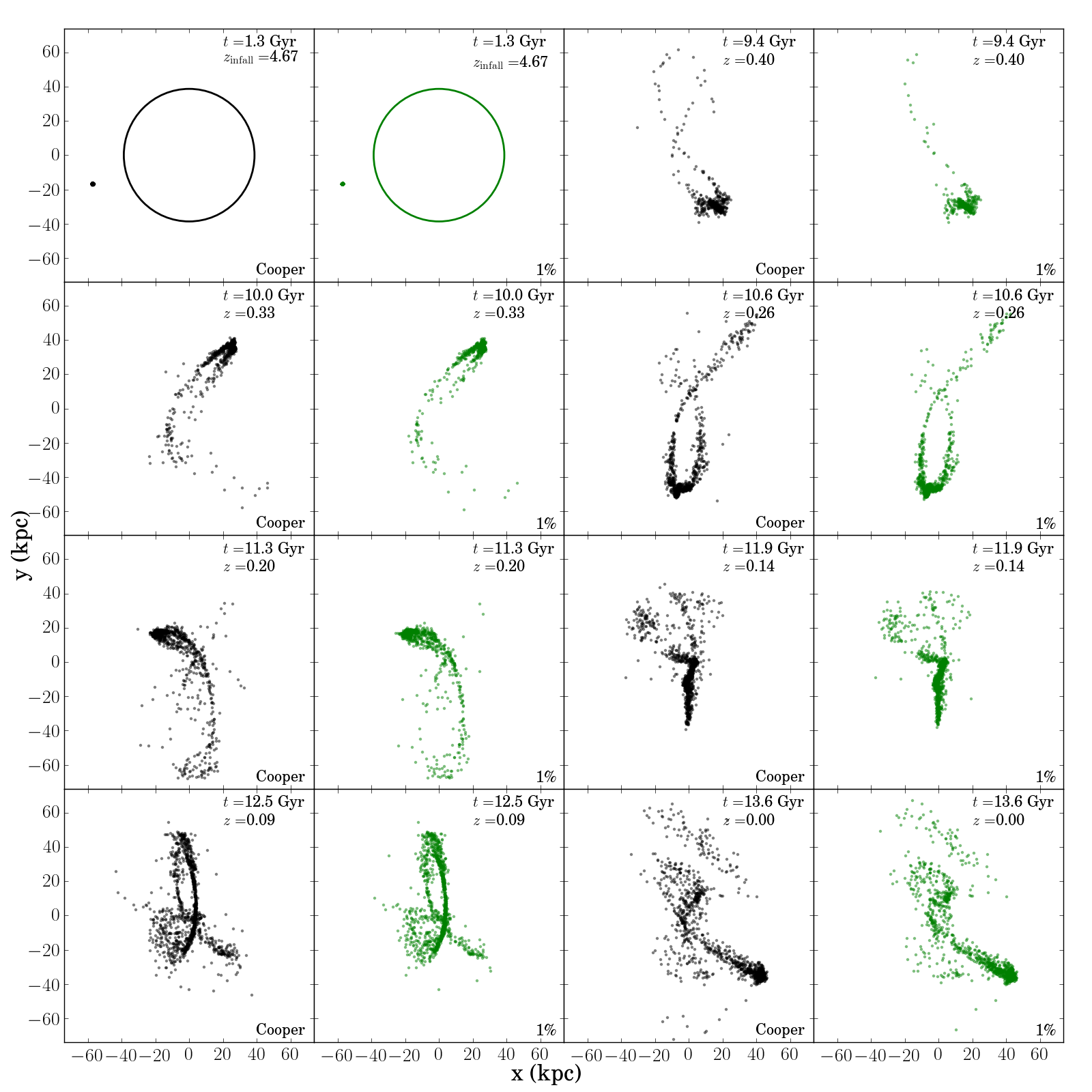}
   \caption{Time-evolution of stream C in both tagging schemes: the one used in Cooper et al. (black) and the 1-percent-at-infall scheme used in this work (green). The dots show the projected locations of the tagged particles relative to the center of the main halo, measured in kpc. The projected axes are the simulation $x$ and $y$ ($y$ is roughly aligned with the major axis of the halo; $x$ lies approximately along the minor axis). Time increases from left to right and then top to bottom, in pairs of panels labeled ``Cooper" and ``1\%" for the two tagging schemes. In the panels labeled ``Cooper," a random selection of 1241 out of the 2056 total tagged particles is plotted for each snapshot so that the same number of particles is plotted for each tagging scheme (the random selection is different in each snapshot). In the two panels on the top left, the circle shows the virial radius of the main Aq-A-2 halo (in the later snapshots the virial radius is larger than the frame).}
   \label{fig:taggingComparisonXY}
 \end{figure*}

 \begin{figure*}
   \centering
 \includegraphics[width=0.95\textwidth]{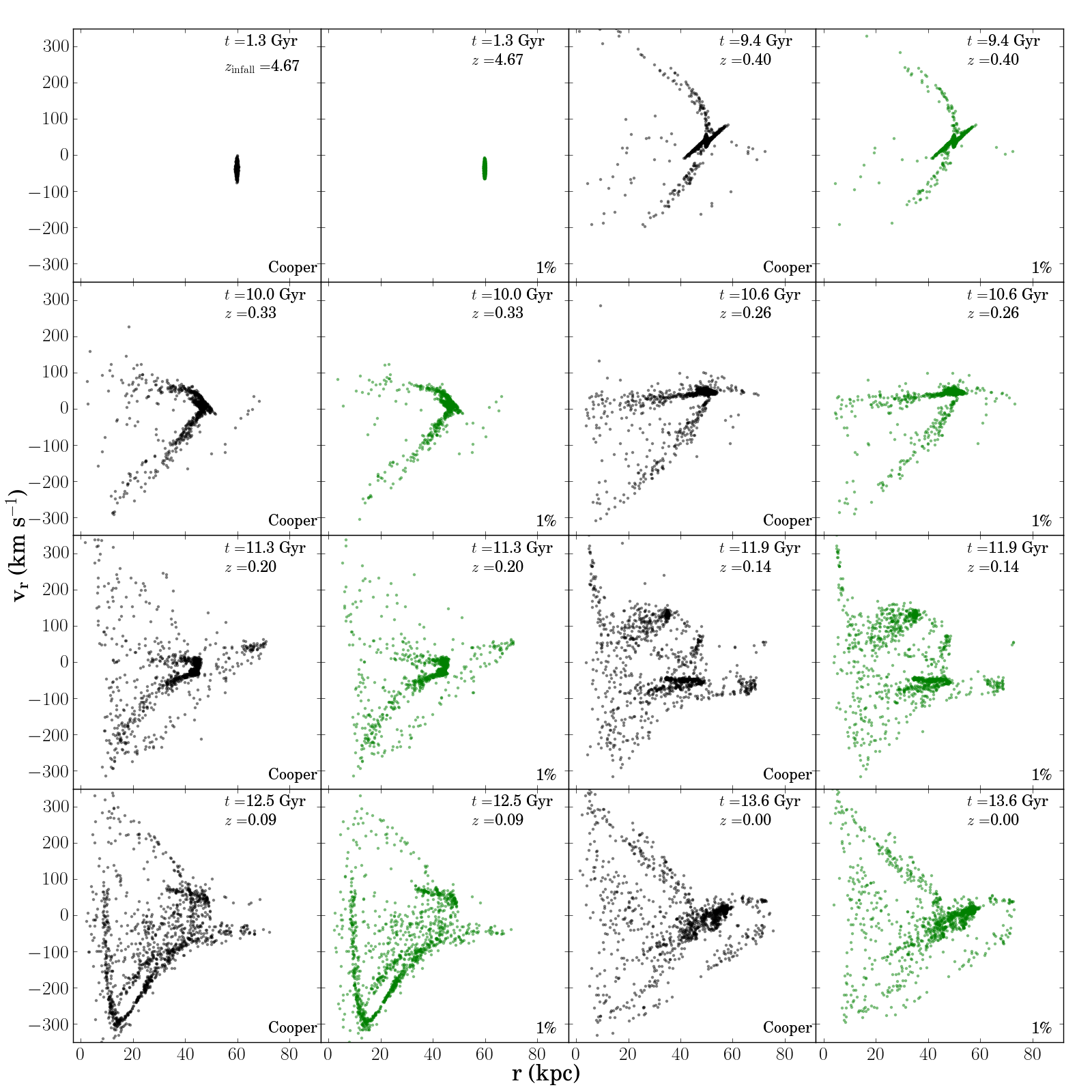}
   \caption{As in Figure \ref{fig:taggingComparisonXY}, but showing the galactocentric radial velocity (in km \unit{s}{-1}) of the tagged particles in stream C as a function of galactocentric distance (in kpc).}
   \label{fig:taggingComparisonRV}
 \end{figure*}

\section{Streams in the tagged Aquarius simulations}
\label{sec:retagging}
Initially we used the scheme described in \citet{2010MNRAS.406..744C} to tag a subset of the dark-matter particles in the Aquarius subhalos as a stellar component, but we found that the assignment of stellar mass to dark matter particles in this scheme can be very lumpy; that is, one dark matter particle can end up with orders of magnitude more stellar mass assigned to it than its neighbor. This makes it difficult to interpret ``gaps'' in the surface brightness of the simulated streams. Because the tagging proceeds throughout the simulation, it is also unclear whether this method evenly samples the phase space of each subhalo. So we also tried a different, much simpler tagging method that assigns stellar mass more evenly to the DM particles.

\subsection{Tagging at infall}
In our new method we also start with the dark-matter-only Aq-A-2 simulation, but tag dark matter particles in subhalos with stellar mass by selecting the most bound 1\% of particles in each subhalo at the snapshot where it is first associated with the main halo in the friends-of-friends tree generated by the halo finder {\sc subfind} \citep{2001MNRAS.328..726S}. We refer to this snapshot as the ``infall'' snapshot. This strategy is similar to that followed by e.g. \citet{DeLucia2008} and \citet{2012ApJ...746..109L}. The amount of associated stellar mass is calculated using the semianalytic model of \citet{2013MNRAS.429..725S}, and is divided evenly between all selected DM particles. For tagged subhalos that are still bound at the present day in the simulation, this produces the luminosity--size relation shown in the lower panel of Figure \ref{fig:lfunc}.  We tried several different tagging percentages up to 10\%, and found that a 1\% tagging fraction most closely matched the size-luminosity relation of the MW and M31 satellites down to the resolution limit of the simulation (where the half-light radius is comparable to a few times the softening length).  The luminosity--velocity dispersion relation is also roughly matched with this tagging fraction.

\begin{figure*}
\begin{center}
\includegraphics[width=\textwidth]{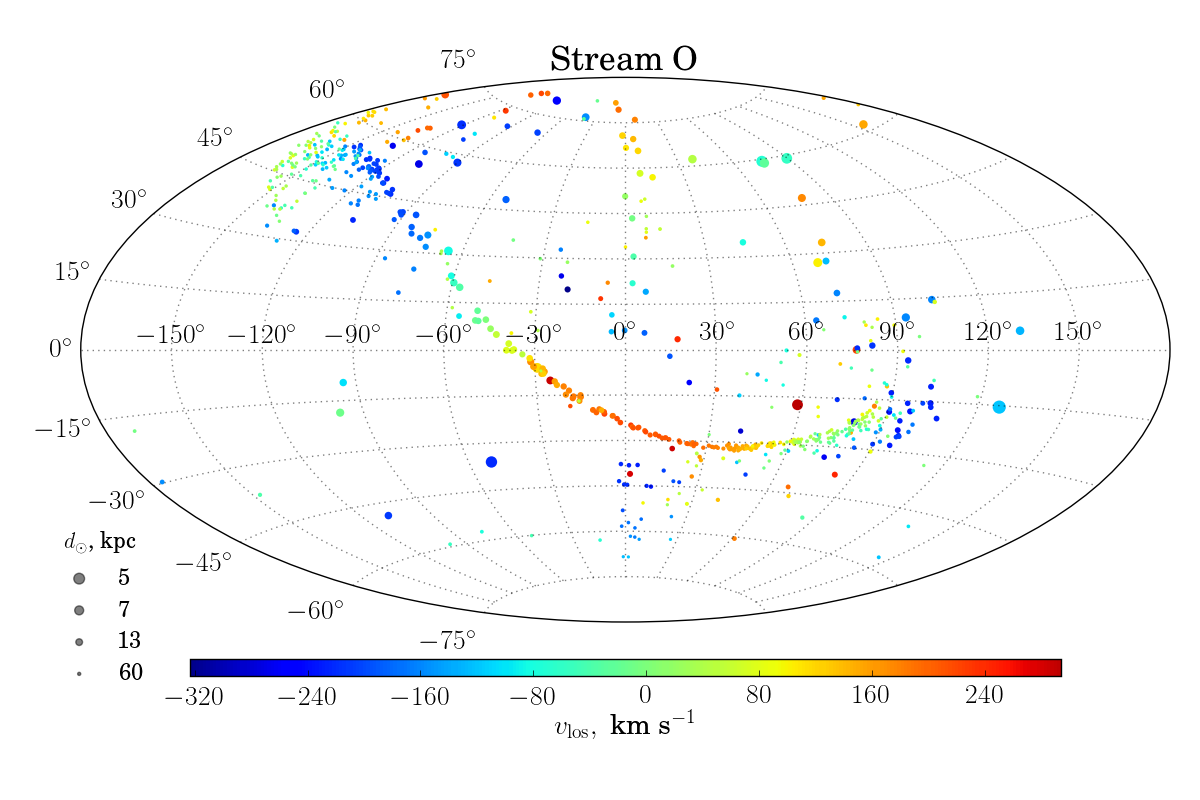}
\caption{View of stream O in ``galactic'' coordinates, for an arbitrary location 8 kpc from the center of the host halo. The size of the points is inversely proportional to distance (closer stars look bigger) while the line-of-sight velocity (negative = approaching) determines color.}
\label{fig:OrphanSky}
\end{center}
\end{figure*}

Tagging methods for representing the building blocks of the stellar halo are to some extent unsatisfactory, since they assume that the phase-space distribution of the stars and dark matter deep within the subhalo potential are identical. Recent work by \citet{2015arXiv150206371L} and \citet{2014ApJ...783...95B} points out discrepancies between the distributions of stars in particle-tagging schemes and those generated by SPH simulations. Some of these discrepancies are due to the way in which tagging proceeds, while others vary with the choice of feedback prescriptions in the subgrid SPH physics, which can either reconcile or exacerbate the differences. However, while neither representation of the streams resulting from tagged DM-only satellites is likely to be correct in detail, we do produce objects that agree with the luminosity-size relationship of the structures that are still bound at present day, which is the best available observational constraint. 

We use a lower tagging percentage than \citet{2015arXiv150206371L},
who found that tagging the most-bound $\sim$5 percent of particles in
subhalos over the course of a dark matter (DM) only simulation
produced the best agreement with hydrodynamic (SPH) simulations. They
hypothesize that a larger tagging region is needed to properly account
for differences in the dark matter particle diffusion rate between the
hydro and DM-only cases (since, for example, the hydro simulations
form a disk that can shock satellites during mergers). However, their
agreement criteria differ from ours; we simply wish to properly
reproduce the observed luminosity-size and luminosity-velocity
relations of the bound MW satellites, while
\citeauthor{2015arXiv150206371L} focus mainly on the structure of the
unbound halo stars and the density profiles of the few largest simulated
satellites. Furthermore we tag at infall rather than over time, so
that the percentages being tagged are not directly
comparable. \citeauthor{2015arXiv150206371L} show that the difference
between tagging 1\% and 5\% of particles is most pronounced in the
massive satellite density profiles, and is much smaller in the unbound
halo structure. On the other hand, \citet{2011MNRAS.418..336L} focused on comparing the
radial profiles of observed stellar halos with those obtained through
tagging and found that tagged fractions of 1--3\% in the
dark-matter-only CLUES simulations could reproduce the observed
stellar halo profile depending on whether the MW, M31, or M33 was
considered, with the lower end of that range being appropriate for the
Milky Way. These authors also looked at CLUES runs including SPH and
found that in these cases a larger tagging fraction (3--5\%) was
required to match the observations, suggesting that matching SPH to
DM-only runs should give different results for the best tagging
fraction than matching either to observations. Additionally, the
choice of tagging fraction will primarily influence the length of the
resulting streams, which (as we will demonstrate in this work) spans a
very wide range, with far more variety than could be produced simply
by manipulating the tagging fraction within the range recommended by
all these studies. We therefore argue that our tagging choices are sufficient for our
purposes and consistent with the studies most similar to our
situation.
\begin{figure*}
  \includegraphics[width=\textwidth]{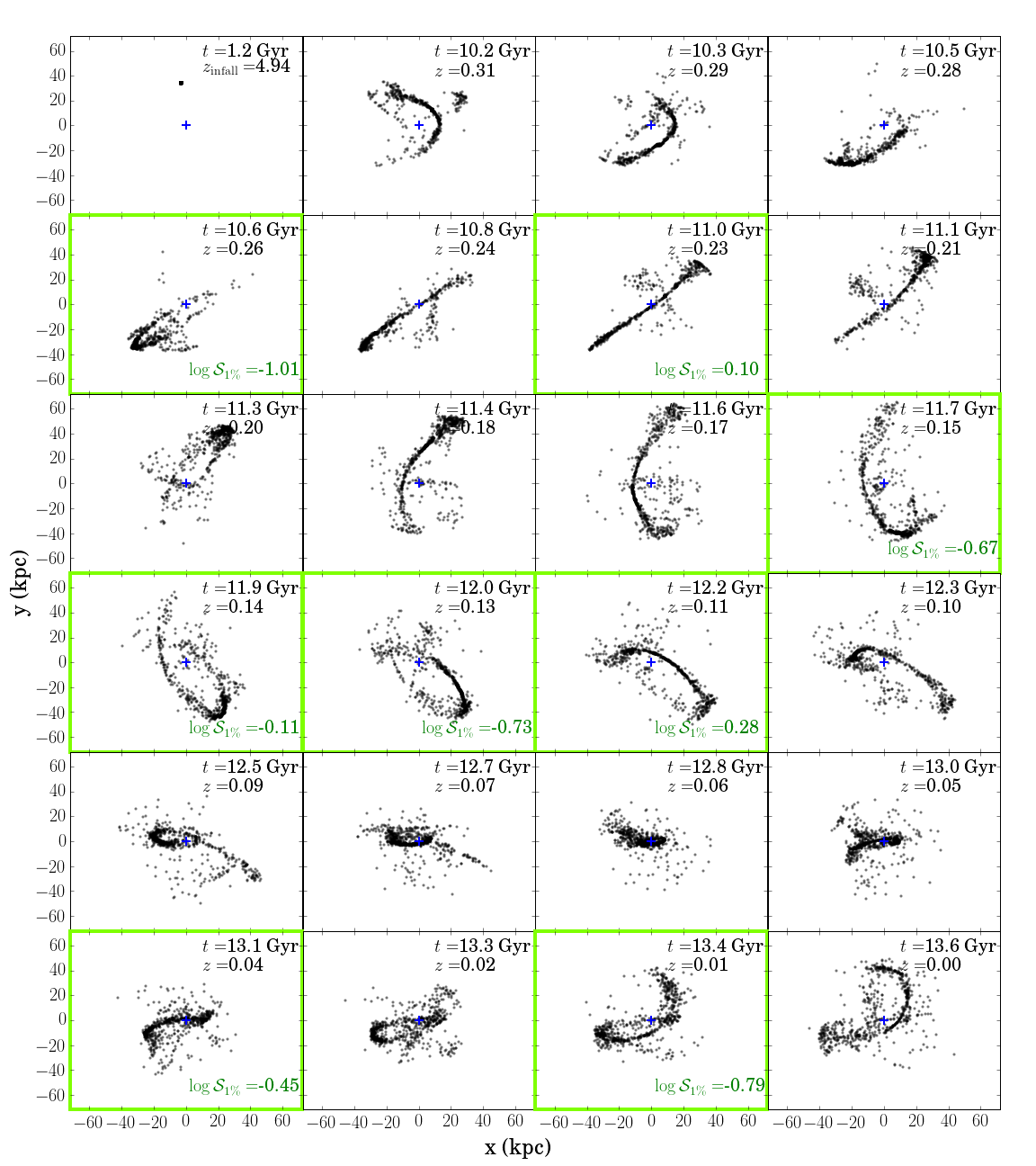}
  \caption{Spatial view of the evolution of stream O with time. The first panel shows the progenitor of the stream at infall. The black points are the ``stars'' (tagged particles) making up the stream. The position axes are centered on the host galaxy center, marked with a blue cross. Frames outlined in green include an interaction with a subhalo at distance less than $r_{1/2}$; in these panels the $\mathcal{S}$ value defined by Equation \eqref{eq:strongEncRatio} is shown in green. }
  \label{fig:OrphanXYsnaps}
\end{figure*}

\begin{figure*}
  \includegraphics[width=\textwidth]{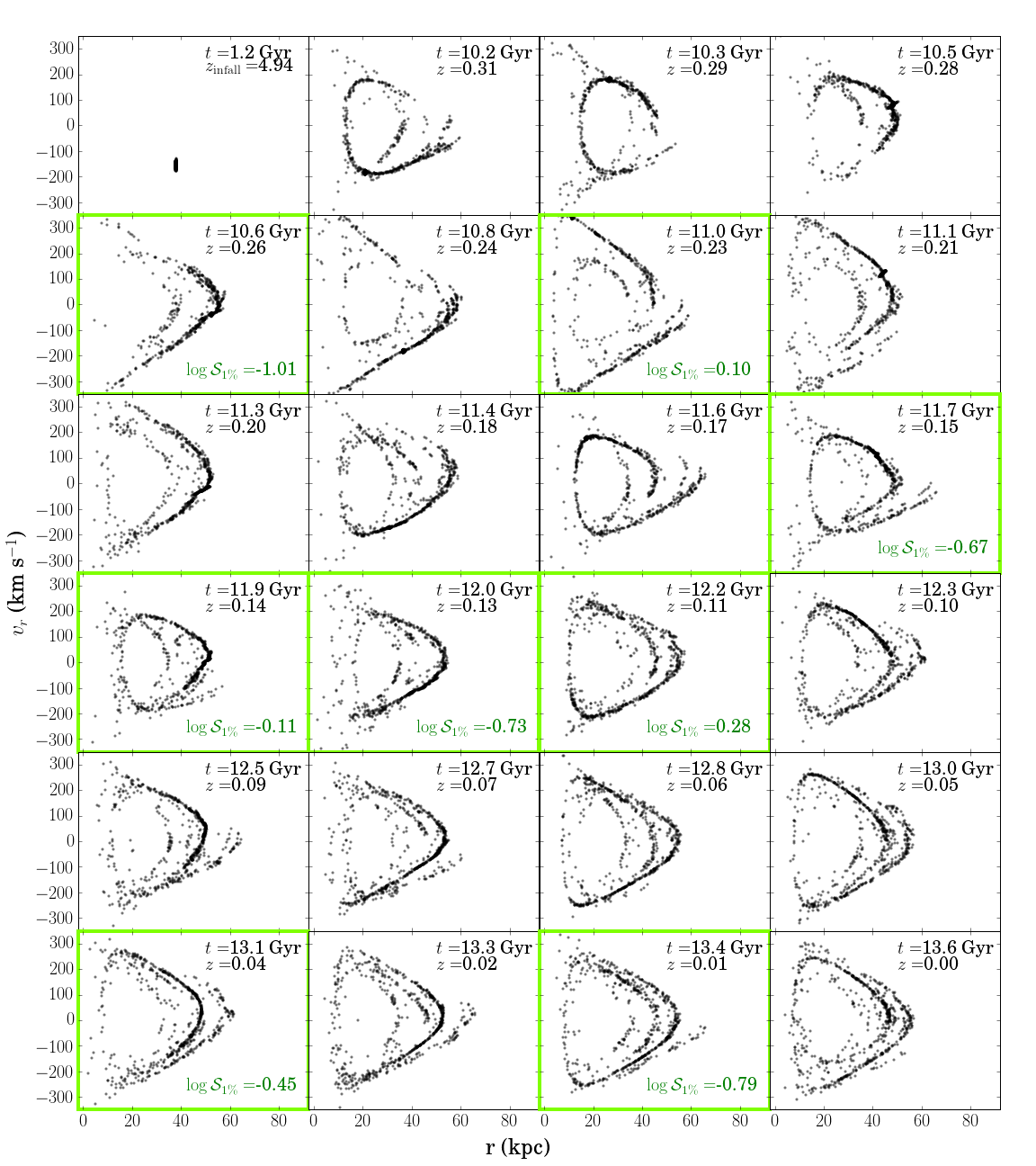} 
\caption{Phase-space view of the evolution of stream O with time. The first panel shows the progenitor of the stream at infall. The black points are the ``stars'' (tagged particles) making up the stream. The radius and radial velocity axes are centered on the host galaxy. As in Figure \ref{fig:OrphanXYsnaps} the green-framed panels mark snapshots where a subhalo is within $r_{1/2}$. }
  \label{fig:OrphanRVRsnaps}
\end{figure*}


\subsection{Comparison to Cooper et al. tagging scheme}

Both our simplified tagging scheme and the method used by \citeauthor{2010MNRAS.406..744C} match the properties of the present-day bound satellites reasonably well, but this agreement does not rule out possible differences in the distributions of tagged particles in the unbound structures. Differences are to be expected, since in \citeauthor{2010MNRAS.406..744C}'s scheme particles are tagged over time, and some dark matter particles could pass into and out of the most-bound region being tagged one or more times. This led us to expect that our single-snapshot tagging would lead to a consistent size of bound structures but without a more diffuse region of tagged particles that were in the subhalo's center at earlier times but not at infall. It was also unclear how much of the apparent lumpiness of the streams in the Cooper tagging scheme had to do with the tagging method, as opposed to structure either in the subhalos themselves or produced in the streams through encounters in the lumpy main halo.

Since we use a different semi-analytic model for our tagging scheme than \citeauthor{2010MNRAS.406..744C}, the particular subhalos chosen for tagging will generally not be the same between the two schemes, especially at lower masses where reionization and/or feedback can suppress star formation in some subhalos but not others. However, most of the highest-mass subhalos should be tagged in both cases and we can use these to compare the appearance of streams in our new scheme with those in \citeauthor{2010MNRAS.406..744C}.  Figures \ref{fig:taggingComparisonXY} and \ref{fig:taggingComparisonRV} show one such stream that illustrates the differences and similarities between the tagging methods. We will refer to this as stream ``C,'' since we use it to {\it C}ompare the two tagging methods. The total stellar mass in this stream is $1.9\times 10^5\ M_{\odot}$ in the Cooper tagging scheme, and $4.6 \times 10^6\ M_{\odot}$ in the new tagging scheme. In both cases the progenitor crosses the virial radius as a very compact object (top left pair of panels) that only begins to tidally disrupt several Gyr after infall. Thus the stream is formed relatively recently even though its progenitor joins the main halo very early. In the figure, we plot the particles tagged using each of the two schemes.

The Cooper version of stream C includes about twice as many tagged
particles as the version in our new tagging scheme (2056 in Cooper
versus 1241 in the new scheme). To highlight differences in the
position and velocity of the tagged particles, we downsample the
Cooper stream by randomly choosing the same number of particles that
are tagged in our new scheme. Once this is done the stream is
virtually identical in the two tagging schemes. Figure
\ref{fig:taggingComparisonXY} shows that in the Cooper scheme there
are a few more outliers and slightly longer tails of tagged particles;
we think these represent particles that were only briefly deeply bound
within the subhalo in the past. However, the substructure within the
stream, visible in both position (Figure
\ref{fig:taggingComparisonXY}) and velocity (Figure
\ref{fig:taggingComparisonRV}), persists regardless of how the
particles are tagged, confirming that it is not an artifact of the
tagging process. As expected, our use of a different SAM also does not
affect our results, since it influences only our calibration of what
percentage of particles to tag as stars, which is set by comparing
with observed bound satellites. The only other way we use the SAM
information is to select streams by stellar mass for further analysis,
but since the simulated satellites are calibrated to match the luminosity function of the
Milky Way satellites \citep{Koposov2008,2013MNRAS.429..725S}, and because we use a wide
range of stellar masses ($10^4$--$10^6\ M_\odot$), we
expect our results not to be sensitive to details of the semi-analytic
model.

\begin{figure*}
\includegraphics[width=\textwidth]{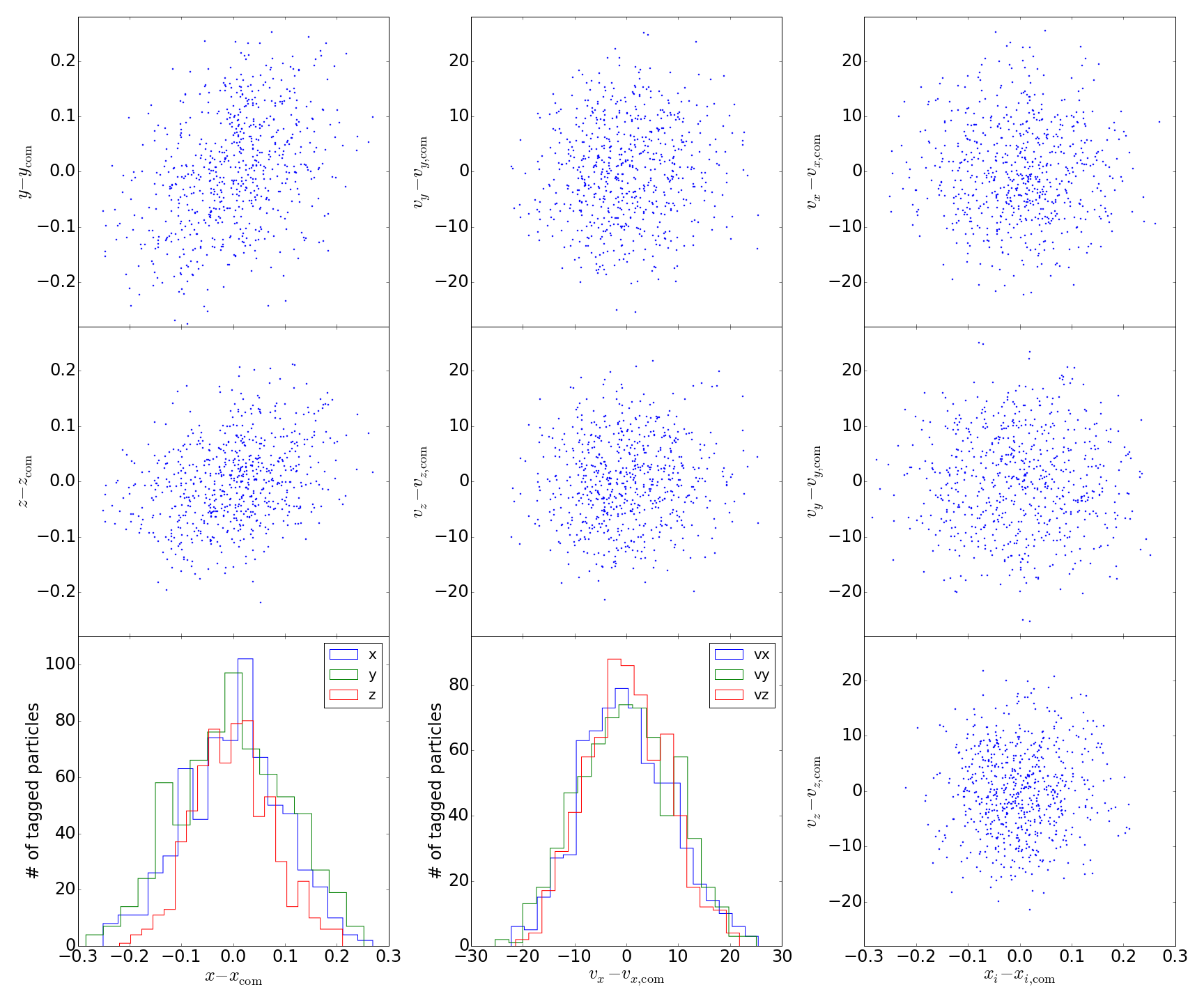}
\caption{Phase-space distribution of the progenitor of stream O at infall. Blue points show the locations of the tagged DM particles relative to the center of mass, in position and velocity. The histograms show the binned one-dimensional distributions of the stars in all 6 phase-space coordinates. All positions are given in kpc; all velocities are in km \unit{s}{-1}.} 
\label{fig:InfallDistribution}
\end{figure*}

\FloatBarrier

\section{Subhalo interactions with an example stream}
\label{sec:exampleStream}

To see how subhalo interactions affect the streams we followed a single stream over time and tracked the number and mass of subhalos that approached the stream within several different threshold distances. We also looked at the evolution of the energy and angular momentum distribution of the stream ``stars'' (tagged DM particles) over time. The stream we tracked resembles the MW's Orphan Stream \citep{2006ApJ...645L..37G,2007ApJ...658..337B}; it has about the same stellar mass ($7.1\times 10^5$ \Msun\ in Cooper, $9.7\times10^5$ \Msun\ in our SAM), and its thickness (a few tens of pc) and angular span (about 1 full wrap) at the end of the simulation are similar. We will refer to it as stream ``O'' (for Orphan). Figure \ref{fig:OrphanSky} shows stream O ``on the sky", for an arbitrary choice of viewpoint 8 kpc from the center of the host halo. The color indicates the line-of-sight velocity, while the size is inversely proportional to the distance (so that closer stars look larger). A total of 857 DM particles in the progenitor subhalo were tagged with stellar mass using the Cooper et al. method, and 673 were tagged using our new method. The dark matter mass of the subhalo at the time of infall was $4.7\times10^8$ \Msun.

\subsection{Initial distribution of the progenitor}
We followed stream O from the point at which its progenitor first becomes a subhalo ($t=1.23$ Gyr) until the end of the simulation ($t=13.58$ Gyr). The stream is well established by $t=9.4$ Gyr. It begins to form at about $t=4$ Gyr, after a few pericenter passages within 20 kpc of the galactic center. As the stream evolves it develops gaps and other density perturbations, features that are consistent with previous studies of the effects of encounters between the stream and dark substructure. In particular we see perturbations both in physical space (Figure \ref{fig:OrphanXYsnaps}) and projected phase space (Figure \ref{fig:OrphanRVRsnaps}) for stream O. However, it is important to determine whether these structures were indeed caused by such encounters, whether they arise from pre-existing structure in the stream progenitor, or whether they result from resolution effects. Thus in Figure \ref{fig:InfallDistribution} we examine the distribution of the tagged star particles in the first snapshot where the progenitor is a subhalo of the main (``Milky Way'') halo, hereafter referred to as the ``infall" snapshot. It appears from the figure that no prominent phase-space correlations exist in the progenitor; the distribution in every coordinate is smooth and roughly Gaussian. If this is so, then the smaller-scale structures apparent in the stream must have been produced after infall. However, some ``holes" are barely distinguishable in the distribution: one example is in the upper left-hand panel of Figure \ref{fig:InfallDistribution} at about (-0.05,0.05). This sort of feature may be due to shot noise in sampling the distribution, and so could also lead to an apparent gap if the noise is correlated in position and velocity. Therefore we wish to confirm that features on this level are consistent with the expected Poisson noise given the number of star particles in the stream.

To confirm quantitatively that the stream starts as a smooth distribution (down to the shot-noise limit) we calculated the mutual information of the distribution in phase space at the infall snapshot. The mutual information compares a distribution $p(\vect{w})$ to the product of its marginals $P(w_i)$:
\begin{equation}
\mathcal{M}[p(\vect{w})] \equiv \int p(w) \log \frac{p(\vect{w})}{\prod_{i=1}^{N_d}P(w_i)} d\vect{w},
\end{equation}
where $N_d$ is the number of dimensions of $\vect{w}$. In our case, $p(\vect{w})$ represents the phase-space density of the progenitor ($\vect{w}$ represents the six-dimensional phase-space position of a particle), normalized by the number of particles so that it integrates to unity. If the distribution is purely a smooth, multivariate Gaussian in all dimensions, with correlation matrix $\mathbf{\rho}$, then we expect $\mathcal{M}$ to be \citep[][Chap. 9]{Kullback59}
\begin{equation}
\sub{\mathcal{M}}{g} = -\half \log |\mathbf{\rho}|.
\end{equation}
In the case of additional substructure within the distribution, $\mathcal{M}>\sub{\mathcal{M}}{g}$. To test whether additional substructure exists, we first calculated $\mathbf{\rho}$ for our distribution and obtained an analytic value for $\sub{\mathcal{M}}{g}$ of 0.15. Then we used the EnBID density estimator \citep{2006MNRAS.373.1293S} to calculate $\mathcal{M}$ for the infall distribution of the 673 star particles in stream O by Monte Carlo integration, obtaining a value of 0.31$\pm 0.02$, about a factor of 2 higher than the analytic estimate. However, previous tests have shown that at small $\mathcal{M}$ this numerical estimator tends to be biased high for samples with a low number of points, so we also constructed 100 random samples of 673 points from a multivariate Gaussian distribution with the same mean and $\mathbf{\rho}$ and estimated $\mathcal{M}$ from these samples, obtaining a value of 0.35$\pm 0.04$, statistically identical to the value for the actual phase-space distribution. Thus we can safely conclude that there is no more structure present in the infall distribution than would be expected for a smooth multivariate Gaussian.

\subsection{Frequency, velocity, and mass of interacting subhalos}
The effect of a subhalo encounter on a stream can be assessed, at zeroth order, by determining whether its distance of closest approach to the stream, $b$, brings it close enough that the potential energy of the nearest stream stars with respect to the subhalo (of mass \sub{M}{sub}) is comparable to the kinetic energy of the stream stars, which move at a velocity \sub{v}{rel} in the rest frame of the subhalo. Thus the condition for a ``strong'' encounter can be expressed as 
\begin{equation}
 \frac{\sub{GM}{sub}}{b} \geq \frac{\sub{v}{rel}^2}{2},
 \label{eq:strongEncounterCondition}
\end{equation}
where $G$ is Newton's constant. One could therefore define a dimensionless ratio, the ``strength" $\mathcal{S}$ of the encounter, characterized by the scaling of Equation \ref{eq:strongEncounterCondition}:
\begin{equation}
\mathcal{S} = \frac{2G\sub{M}{sub}}{\sub{v}{rel}^2 b},
\end{equation}
so that strong encounters have $\mathcal{S} \gtrsim 1$. However, this criterion is quite conservative compared to what would actually be needed to induce a local gap in a stream: an $\mathcal{S} \sim 1$ interaction could for example reverse the direction of motion for stars in the stream, where what is actually needed to produce a gap in a coherent stream is to change the velocities of a small group of stream stars by an amount comparable to the stellar velocity dispersion, which is usually a few percent of $\sub{v}{rel}$ (i.e. a few km/s). This likewise corresponds to a few-percent change in the kinetic energy of the stars in the stream, so we choose to rescale $\mathcal{S}$ by the fractional velocity change $\eta$,
\begin{equation}
\mathcal{S}_{\eta} = \frac{2G\sub{M}{sub}}{\eta \sub{v}{rel}^2 b},
\end{equation}
and set $\eta=0.01$ so that an $\mathcal{S} \sim 1$ interaction changes the velocity of the interacting stream stars by one percent. Thus we define the quantity $\mathcal{S}_{1\%}$, using
\begin{equation}
\mathcal{S}_{1\%} = 100 \cdot \frac{2G\sub{M}{sub}}{ \sub{v}{rel}^2 b},
\label{eq:strongEncRatio}
\end{equation}
as a way to tell how likely an interaction is to open a gap in the stream.

By following stream O through the simulation, we can track the encounters it experiences and ask how many of them fit this basic criterion. To determine the type and rate of interactions between stream O and subhalos, we counted the number of subhalos in each snapshot for which the most-bound particle in the subhalo passed within $b=1$, 2, and 5 kpc of any star particle in the stream. We also counted how many subhalos passed within one or two times their half-mass radius, $r_{1/2}$, reflecting the region of influence in which we expect a subhalo to significantly alter the orbits of the stream stars. We tracked the identification numbers of the subhalos interacting with the stream to determine whether there were any repeat encounters with the same subhalo in different snapshots.

\begin{figure}
 \includegraphics[width=0.5\textwidth]{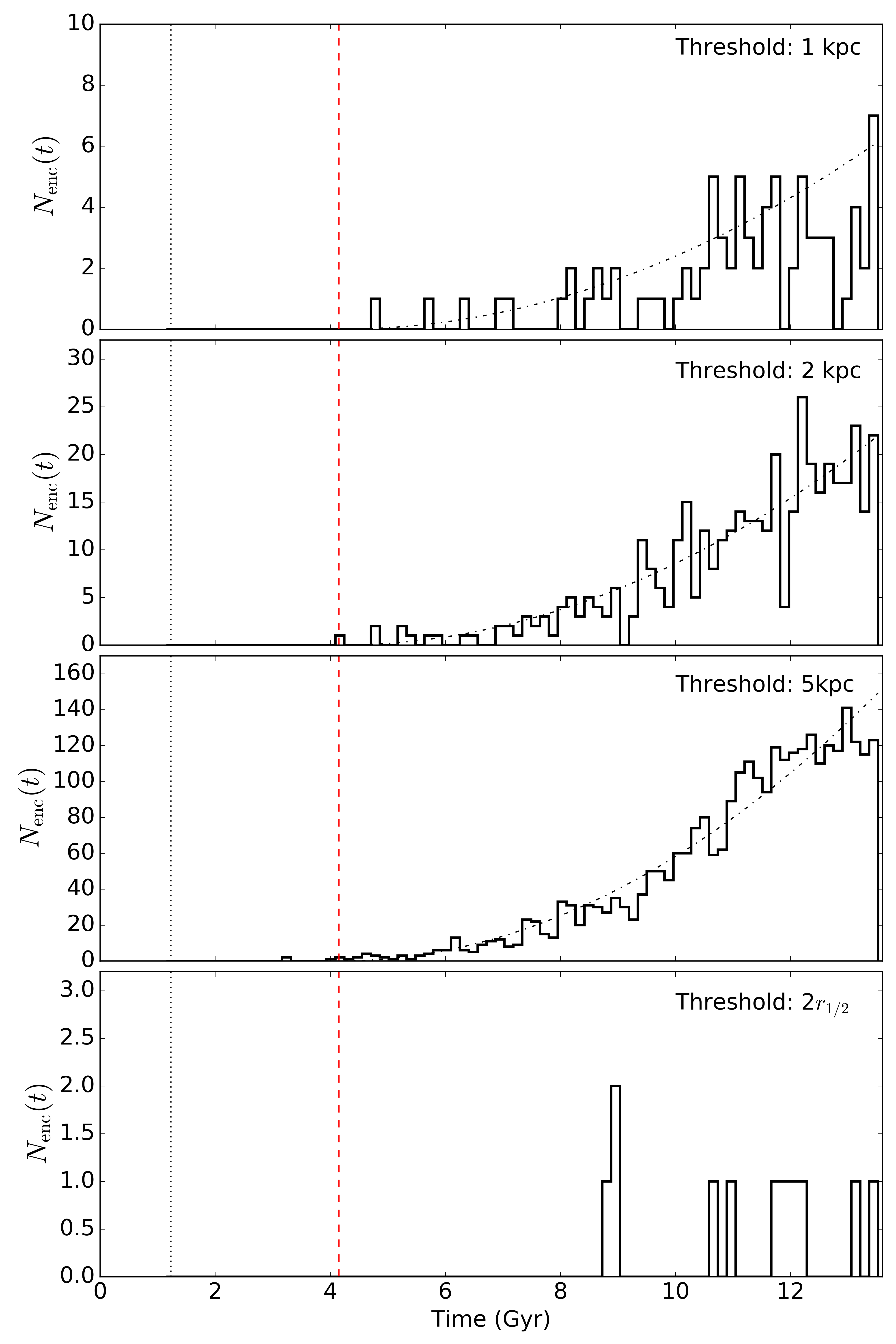}
\caption{Number of interactions between subhalos and any tagged star particle in stream O, as a function of time, for different interaction thresholds $b$. Only interactions after infall ($t=1.23$ Gyr; dotted vertical line) are shown. The time when the stream starts to form (identified as described in Section \ref{subsec:lengthAge}) is shown as a red dashed vertical line. In the top three panels, the dot-dashed line indicates the $t^2$ scaling posited by YJH, roughly fit by eye.}
\label{fig:OrphanIntxAll}
\end{figure}

\begin{figure}
  \includegraphics[width=0.5\textwidth]{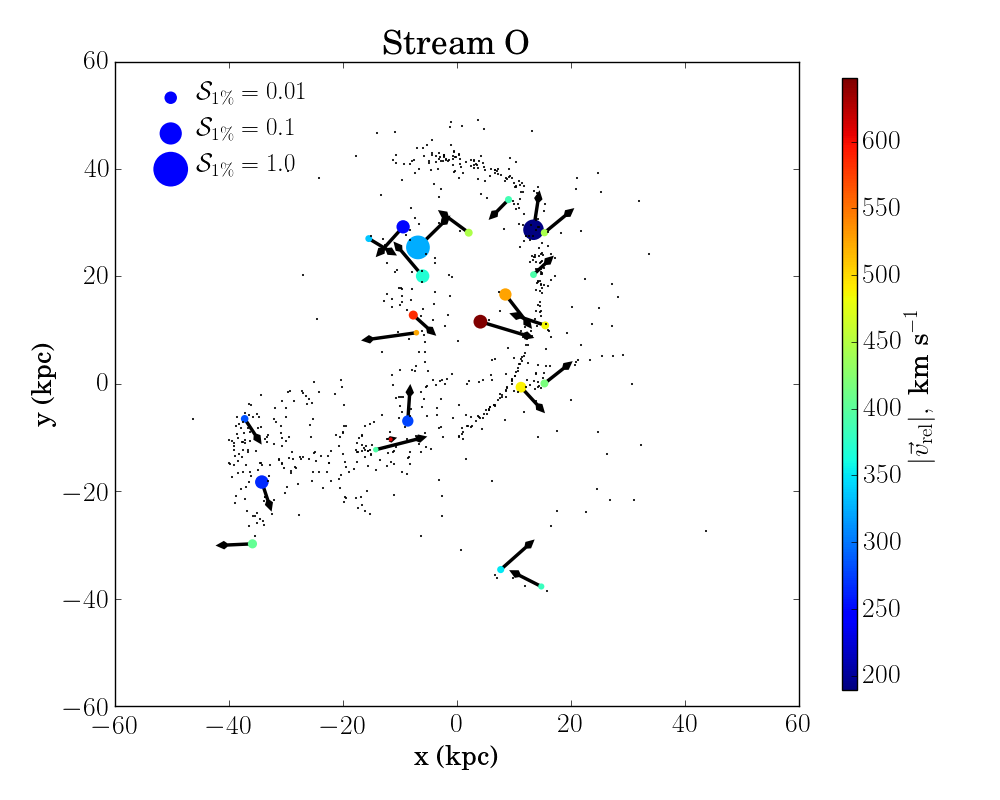}
\caption{Stars in stream O (black) and all subhalos within 2 kpc at present day (colored points with arrows). Each subhalo's velocity vector is shown. Subhalos are colored according to absolute relative velocity, and their size is proportional to the ratio given by Equation \eqref{eq:strongEncRatio} with $b$ the closest distance to any star in the stream. Values of this ratio closer to 1 imply stronger encounters. }
\label{fig:closeEncounters}
\end{figure}

The number of subhalos interacting with stream O in each snapshot, at the various distance thresholds, are plotted in Figure \ref{fig:OrphanIntxAll}. The $r_{1/2}$ threshold is not shown as there is only one interaction within this threshold over the course of the simulation, at 12 Gyr with a $10^7 M_\odot$ subhalo. This is not too surprising given that for most of the interacting subhalos, $r_{1/2}$ is much smaller than 1 kpc. As expected, the interaction rate increases with time as the stream increases in length, and roughly follows the $t^2$ scaling predicted by YJH (dot-dashed lines), beginning at the start of the formation of the stream at $\sim$6 Gyr. The number of encounters starts to fall beneath this scaling at late times, especially in the 2 and 5 kpc panels. We think this is due to particle resolution effects, which we discuss in more detail in Section \ref{subsec:resolution}. We do not see any pattern in the number of events with time for the $2r_{1/2}$ threshold (bottom panel), which is consistent with noise. Particle and time resolution probably combine to cause us to miss many of these encounters, so based on this result we decided to analyze encounters for the subhalo-dependent thresholds only in aggregate, and do not calculate rates for individual streams.

\begin{figure}
   \centering
 \includegraphics[width=0.45\textwidth]{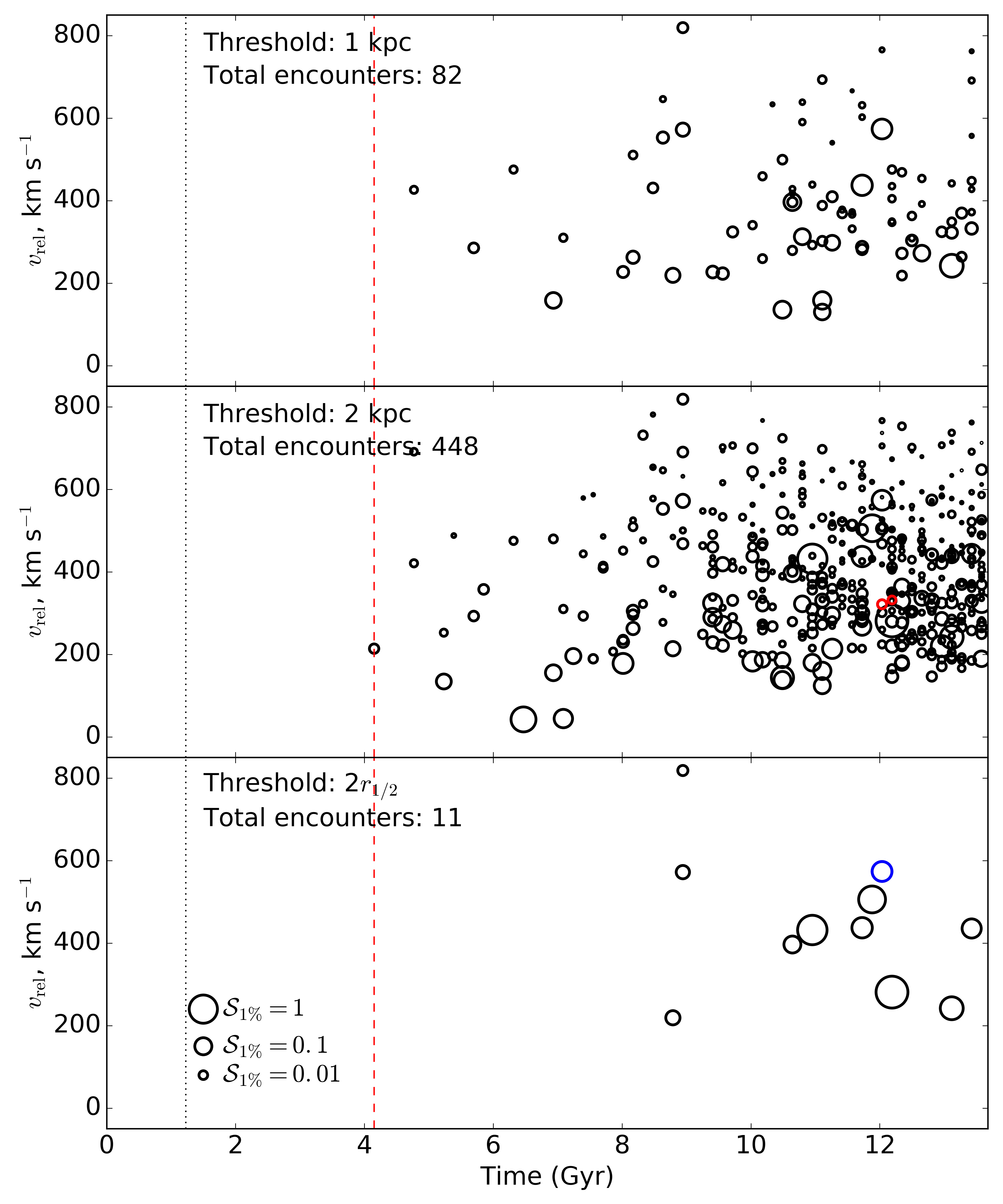}
  \caption{Encounters of stream O with subhalos as a function of time and speed relative to the stream, for the thresholds shown in Figure \ref{fig:speedDist}. Subhalos that have more than one encounter with the stream are shown in red. One-time encounters are shown in black; the encounter shown in blue in the bottom panel is within $r_{1/2}$. Only encounters after infall are shown. As in Figure \ref{fig:closeEncounters}, the size of the circle is proportional to the value of the ratio $\mathcal{S}_{1\%}$ in Equation \eqref{eq:strongEncRatio}. The infall time is shown as a vertical dotted line; the time when the stream starts to form (identified as described in Section \ref{subsec:lengthAge}) is shown as a red dashed vertical line.}
   \label{fig:velmasstime}
 \end{figure}

Figure \ref{fig:OrphanIntxAll} illustrates clearly that in any given snapshot after stream O is formed, it is interacting with many subhalos: up to 6 at a time within 1 kpc and up to 26 at a time within 2 kpc. Figure \ref{fig:closeEncounters} shows the stream with all subhalos within 2 kpc at present day. The velocities of the subhalos do not appear to be correlated with the velocity of the stream, implying that these subhalos did not simply fall in at the same time as the stream progenitor on similar orbits and follow the stream through space, but are passing though the stream with many different impact angles. This is also illustrated when we examine the interaction data for repeat encounters: one would expect these to occur frequently if the subhalos responsible for the interactions fell in with the stream along similar orbits. However, Figure \ref{fig:velmasstime} shows that all but one of the $\sim450$ interactions inside 2 kpc are one-time events.

We recorded the relative velocities of the interacting subhalos with respect to the stream stars within the threshold distance, in order to rank encounters by their strength and determine whether the field of perturbers is truly isotropic. For each subhalo interaction we identified the \sub{N}{close} stars within the threshold distance and calculated the average vector velocity $\vect{\bar{v}}$ of those stars component-by-component:
\begin{equation}
\vect{\bar{v}} =  \frac{1}{\sub{N}{close}} \sum_{j=1}^{\sub{N}{close}} \vect{v}_{j}.
\end{equation}
Then we determined the relative velocity by subtracting this from the velocity of the subhalo:
\begin{equation}
\sub{\vect{v}}{rel} = \sub{\vect{v}}{sub} - \vect{\bar{v}}.
\end{equation}
The vector relative velocity was tabulated for every interaction in every snapshot.

\begin{figure*}[!p]
\begin{center}
\includegraphics[width=\textwidth]{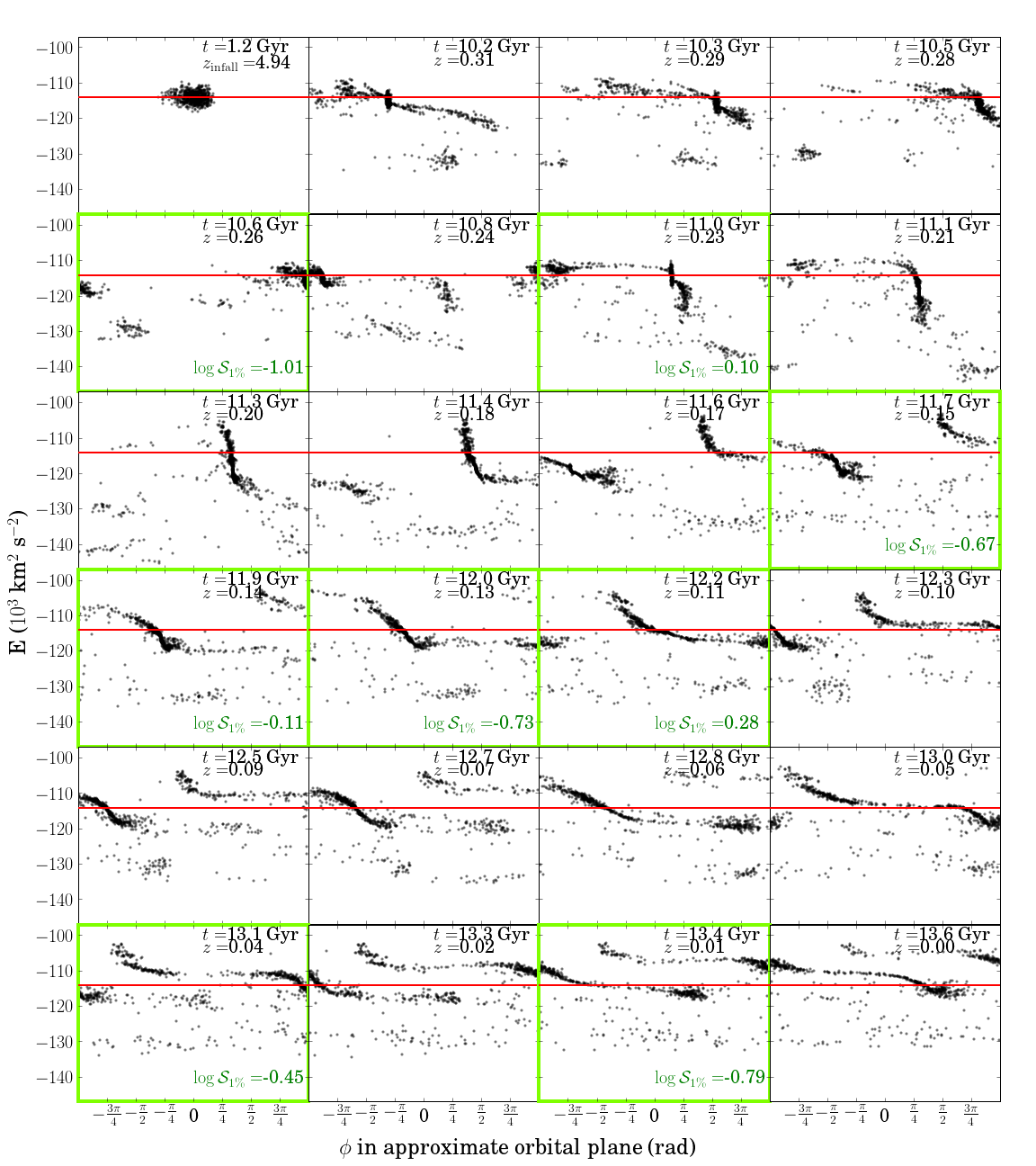}
\caption{Energy versus phase angle for particles in stream O as a function of time. The red line marks the mean energy at infall (upper left panel). As in Figures \ref{fig:OrphanXYsnaps} and \ref{fig:OrphanRVRsnaps} the green frames mark snapshots with interactions at $b<r_{1/2}$.}
\label{fig:energyOrphan}
\end{center}
\end{figure*}

Figure \ref{fig:velmasstime} illustrates how, although many subhalo interactions do occur, most of them are likely too weak to substantially alter the stream. The size of the symbols in the figure correspond to the strength $\mathcal{S}_{1\%}$ of the encounter  as defined in Equation \ref{eq:strongEncRatio}. The relative velocity \sub{v}{rel} is determined using particles within the impact parameter $b$, so $\mathcal{S}_{1\%}$ can vary slightly for different impact parameters. For the fixed distance thresholds, the mass distribution of subhalos in encounters reflects the steepness of the halo mass function in general, so one would expect most of the encounters to be very weak but also very frequent. Figure \ref{fig:velmasstime} shows that numerous relatively weak encounters are generally the rule over the lifetime of the stream. Only a handful of subhalos come within $2r_{1/2}$ of the stream over its lifetime, mostly more massive subhalos since these have larger half-mass radii. As shown in the bottom panel, selecting encounters with $b\leq 2r_{1/2}$ is similar (though not identical) to selecting encounters with large $\mathcal{S}_{1\%}$, which are those most likely to open a gap in the stream.


\subsection{Energy and angular momentum evolution}
\label{subsec:EandL}
The encounters between stream O and subhalos also leave an imprint on its energy and angular momentum distribution. Based on the results of YJH and the work of Carlberg and collaborators \citep{Carlberg2009,CarlbergConf2012,Carlberg2012,2013ApJ...775...90C,2014ApJ...788..181N,2015ApJ...800..133C,2015ApJ...803...75N,2015ApJ...808...15C}, we expect that the strongest subhalo encounters will rearrange part of the stream's energy-angular-momentum distribution. To investigate this, we followed the energy over time (including a computation of the potential energy directly from the N-body snapshot) as a function of the relative phase of the stream particles, as shown in Figure \ref{fig:energyOrphan}. To compute the phase angle $\phi$, we rotate the stream into the plane perpendicular to its mean angular momentum $L$, then compute the angle $\tan \phi \equiv y'/x'$, where $x'$, $y'$ are coordinates in the rotated frame. The angular momentum direction is not constant in time, as one would expect given that the potential is triaxial, and in fact it changes by of order 90 degrees over the course of the simulation. Since this change is large, we use the angular momentum of the stream in a given snapshot to compute the phase angle, so the orbital plane in each panel of the figure is different. The average energy at infall (red horizontal line) is roughly preserved as a function of time.

At $t$=10.2 Gyr long tidal tails have already formed around the remaining bound structure, and show signs of disturbance especially along the ends. There is also some structure that appears disconnected from the main energy distribution, around $E \sim -1.3\times 10^5\ (\textrm{km}\ \unit{s}{-1})^2$; this material was energetically detatched from the main body at around 8.2-8.3 Gyr. From about 11--11.5 Gyr the tails appear to rotate vertically in this view; Figure \ref{fig:OrphanXYsnaps} shows that this is because most of the stream is near apocenter. 

The closest encounter (within $r_{1/2}$) is with a $10^5 M_\odot$ subhalo at 11.7 Gyr. The time-resolution of the snapshots prevents us from drawing a direct connection between either of these encounters and specific changes in the energy distribution of the stream stars, but the types of discontinuous features seen in the figure are reminiscent of those found by YJH in their simulations of a globular-cluster-like stream. Suggestively, the trailing tail in the 11.9 Gyr panel is sharply truncated. None of these so-called ``gaps'' are entirely devoid of star particles, but manifest rather as low-density regions. As time goes on the stream continues to lengthen, but displays significant variations in density that also widen with time. These gaps were likely created by subhalo encounters, but although we can identify snapshots where some of the closest encounters occur, the disturbances are so small in energy (recall $\mathcal{S}_{1\%}=1$ is a one-percent change in kinetic energy) that they are hard to identify in energy space in the same snapshot. Because the effects appear only after a delay while the gap grows in size, we cannot with certainty connect a particular subhalo encounter with a specific stream perturbation at the time-resolution available in this simulation.

\FloatBarrier

\begin{figure*}
\begin{center}
\includegraphics[width=\textwidth]{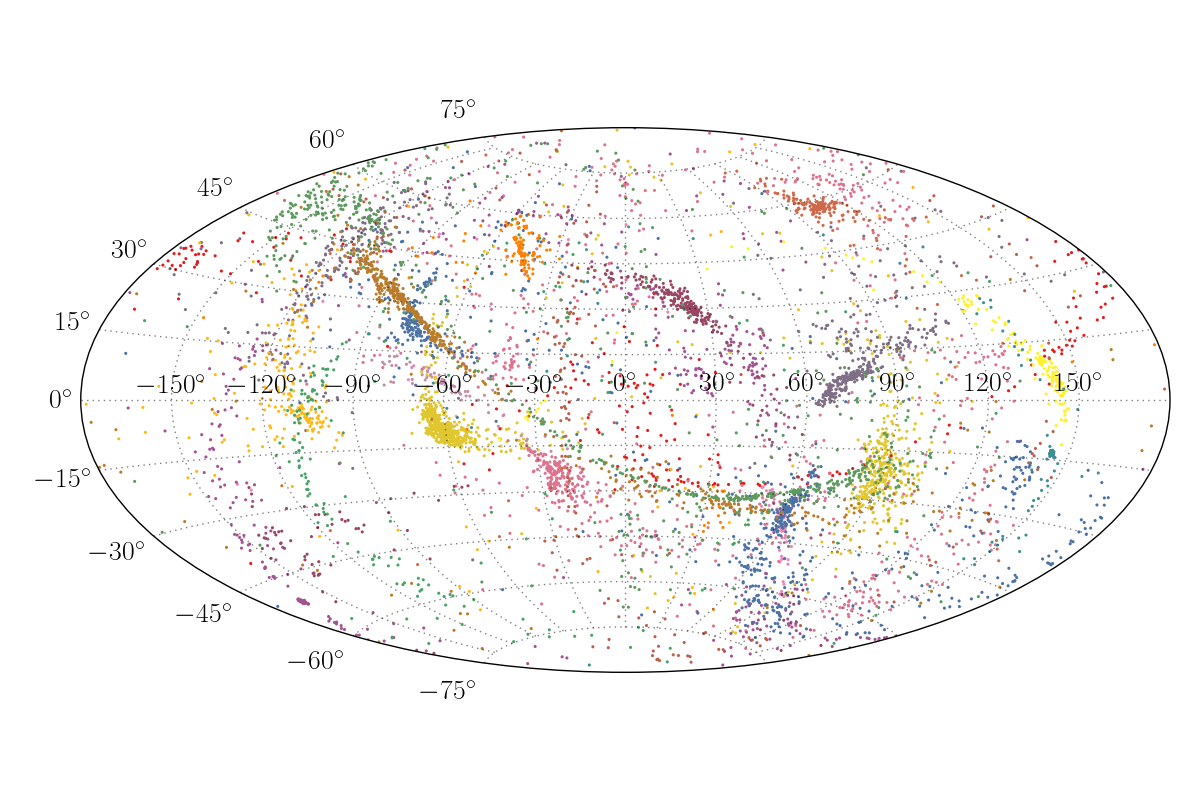}
\caption{View of the set of thin streams in ``galactic" coordinates, for an arbitrary location 8 kpc from the center of the host halo. Each stream is a different color; some look thin from this vantage point and some do not.}
\label{fig:AllSky}
\end{center}
\end{figure*}

\section{Statistics of stream-subhalo interactions}
\label{sec:intxnStats}

We also considered a larger subset of streams in the Aq-A-2 halo that appear thin at the present day based on the retagging of star particles described in Section \ref{sec:retagging}. First we selected all streams with stellar mass between $10^4$ and $10^6\ \Msun$. Then we visually inspected the present-day configuration of these streams and picked ones that were thin and not well-mixed in position and velocity space. These criteria resulted in a total sample of 18 streams, shown in sky projection in Figure \ref{fig:AllSky}. For convenience we labeled these streams with letters from A to R, with stream O assigned as before. Stream C, which was used in Section \ref{sec:retagging} to compare the different tagging schemes, is above the high end of this mass range and so is not included in the sample: we wanted a very well-resolved stream for the comparison of the methods, but the last panel of Figure \ref{fig:taggingComparisonXY} shows that this stream is fairly thick and disorganized at present day.

\begin{figure*}
\begin{center}
\begin{tabular}{cc}
\includegraphics[width=0.45\textwidth]{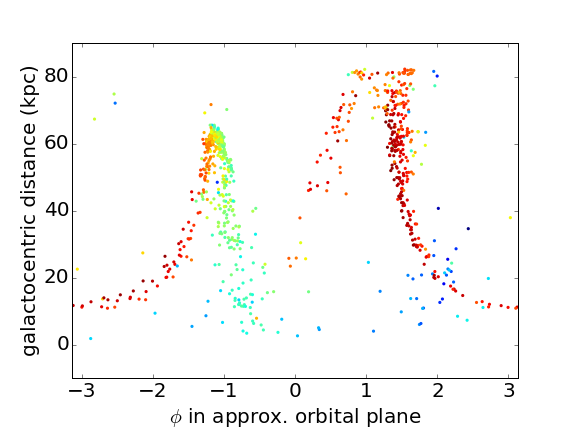} & \includegraphics[width=0.45\textwidth]{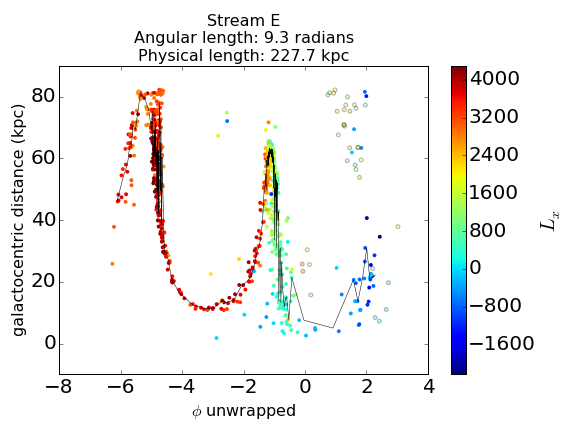}
\end{tabular}
\caption{Left: stream E from the sample, rotated into the plane perpendicular to its mean angular momentum vector. In both panels, the colors show the component of angular momentum (in the non-rotated frame) used to unwrap the stream. Right panel: the same stream unwrapped in the mean plane with outliers shown as open circled points. The black line is used to measure the physical length of the stream as described in the text.}
\label{fig:unwrapping}
\end{center}
\end{figure*}

\subsection{Measuring the lengths and ages of streams}
\label{subsec:lengthAge}


Since the streams have a wide variety of orbits and ages, we wished to normalize the interaction statistics by the length of each stream at the present day and by each stream's age. To measure stream length, we first calculate the phase angle $\phi$ as described in Section \ref{subsec:EandL}. An example is shown in the left panel of Figure \ref{fig:unwrapping}. The stream must then be ``unwrapped" in $\phi$, which we accomplish by adding and/or subtracting multiples of $2\pi$ for subsets of the stream particles so that the components of the angular momentum $\mathbf{L}$ (in the unrotated frame) vary continuously, since overlapping wraps tend to have very different values of at least one angular momentum component (right panel of Figure \ref{fig:unwrapping}). To track the continuous variation of $\vect{L}$ we pick the component in the unrotated frame which has the widest range of values; in the example shown in Figure \ref{fig:unwrapping} this is $L_x$ but it varies from stream to stream. Stream particles that are outliers in angular momentum and cannot be assigned unambiguously to a particular wrap are masked (light circled points in the right panel of Figure \ref{fig:unwrapping}). The angular span is derived by taking the range of the unwrapped angle without the masked particles, and the physical length is computed by binning small numbers of consecutive particles in the unwrapped $\phi$ coordinate, computing their average radius $\bar{r}_i$ and angular range $\delta\phi_i$, and calculating the Riemann approximation
\begin{equation}
\ell_s = \sum_{i=1}^{\sub{N}{bins}} \bar{r}_i \delta\phi_i.
\end{equation}
The number of particles per bin ranges from 1 to 10 depending on the number of particles in a particular stream, and is adjusted to give a relatively smooth approximation to the stream everywhere (thin black line in the right panel of Figure \ref{fig:unwrapping}).

As shown in the case of Stream O, there is usually a delay between the time that the parent halo becomes a satellite of the main galaxy and the time when a stream begins to form. We therefore determined for each stream in our sample the point when the stream started to form and used this to calculate the stream's age. To do so we compared the largest and smallest eigenvalues of the covariance matrix of the positions of the tagged star particles associated with each stream as a function of time. Before the stream starts to form, the stars are all still bound to the satellite so all three eigenvalues should be similar, while once the satellite begins to tidally disrupt, one eigenvalue (the one along the spreading direction) will become much larger. In practice, the ratio of largest to smallest eigenvalue is between about 1.5 and 4 for all our streams while in the satellite phase, and increases rapidly once the stream starts to form. As the stream wraps around the galaxy the ratio of eigenvalues begins to oscillate, and can decrease again as the stream becomes phase mixed, but the initial increase is diagnostic in every stream we examined. Based on this behavior we identified the snapshot where the ratio of eigenvalues is first larger than 5 as the start of stream formation, found the corresponding formation time $\sub{t}{form}$, and computed the age as 
\begin{equation}
t_s = 13.6 \textrm{Gyr} - \sub{t}{form}.
\label{eq:age}
\end{equation}
Encounters before $\sub{t}{form}$ are very rare (less than one per stream) thanks to the small cross-section of the bound progenitor, so we consider the number of interactions since infall to be equivalent to the number since formation.

\begin{figure*}
\includegraphics[width=\textwidth]{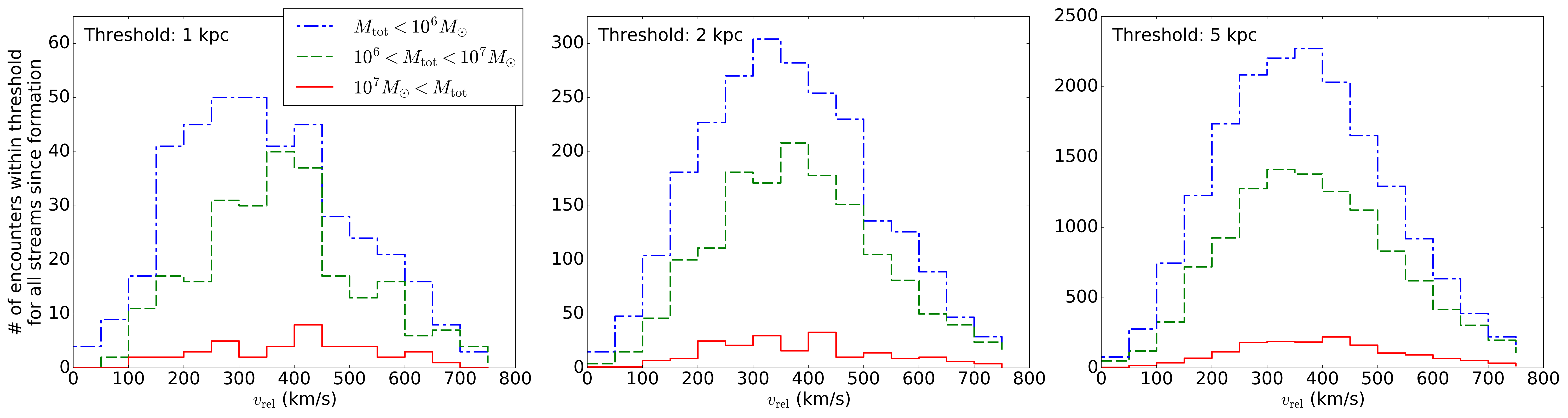}
\caption{Distribution of average relative velocities between subhalos encountering any stream in the sample and the stars within the encounter radius, for three different choices of encounter radius (from left to right): 1 kpc, 2 kpc, and 5 kpc. The colors show different mass bins. Only encounters after the start of stream formation (identified as discussed in Section \ref{subsec:lengthAge}) are counted.}
\label{fig:speedDist}
\end{figure*}

\begin{figure*}
\begin{center}
\includegraphics[width=\textwidth]{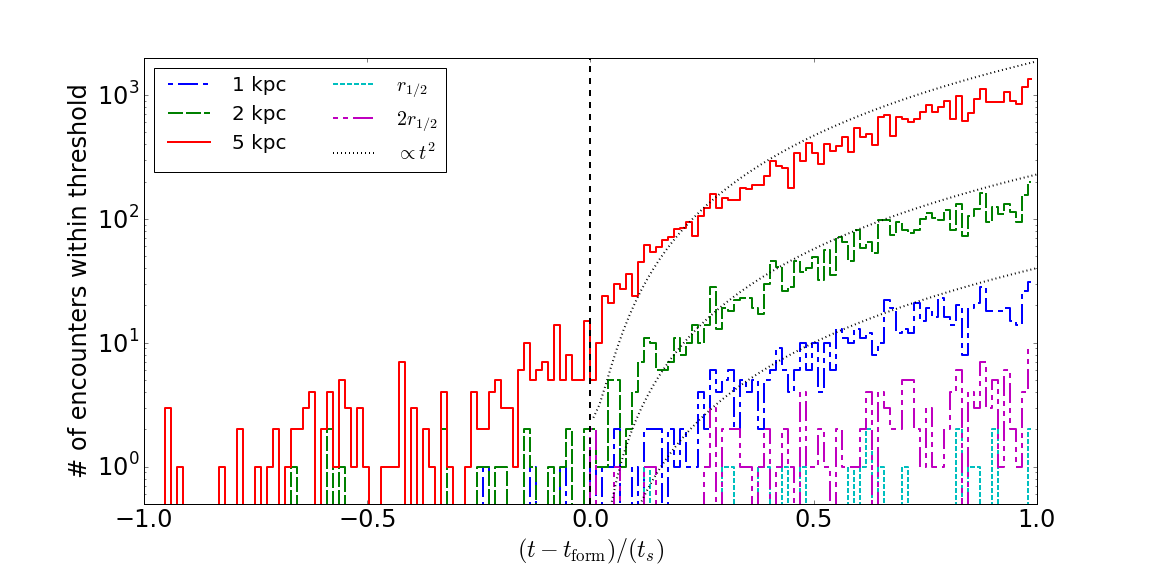}
\caption{Total number of encounters per snapshot for all 18 streams tracked, for the five different thresholds considered. The encounters are stacked relative to the formation time of each stream $t_{\mathrm{form}}$ and timescales are normalized by the age of each stream $t_s$ (Section \ref{subsec:lengthAge}). The dotted gray lines show the expected time-evolution of the number of encounters within a fixed threshold postulated in YJH, $N_{\mathrm{enc}} \propto t^2$, fit to the scaled time range (0,0.25) for the three fixed distance thresholds.}
\label{fig:enc-with-time-all}
\end{center}
\end{figure*}

\subsection{Relative velocities of interactions}
\label{subsec:vrel}

Figure \ref{fig:speedDist} shows the distribution of encounter speeds for three bins in subhalo mass. We used a two-sample Kolmogorov-Smirnov (K-S) test to determine whether different-mass subhalos are drawn from the same speed distribution by comparing pairs of distributions in different mass bins for the same threshold.  For the mass-independent thresholds shown in the figure, the K-S test generally detects a difference between the distributions in different mass bins where there are enough subhalos in the sample to do so, and the certainty with which one can conclude the distributions are different increases with the sample size. For the mass-dependent thresholds (not shown) the K-S test says all the distributions in different mass ranges are indistinguishable ($p\sim 0.86-0.92$), probably due to the small sample size. 

To investigate differences in the relative speed distribution for different subhalo masses, we fit a Maxwell-Boltzmann (M-B) distribution to the CDF of relative speeds for each mass range shown in Figure \ref{fig:speedDist}. Although the M-B distribution does not describe the overall velocity distribution of dark matter in Aq-A-2 very well \citep{2009MNRAS.395..797V}, it is a fairly good fit to the relative speed distributions. The fits to different mass ranges result in velocity dispersions that systematically differ by about 10-15\%, with the lowest values (220--235 km \unit{s}{-1}) coming from the lowest mass range and the highest values (235--255 km \unit{s}{-1}) obtained for the highest mass range. 

The broad speed distributions we find for the interacting subhalos are roughly in agreement with the reasoning outlined in \citet{Yoon2011} that the interaction speed distribution should be Maxwellian with characteristic width $\sqrt{2}\sigma$, where $\sigma$ is the velocity dispersion of the subhalos. Fitting a M-B distribution to the speed distribution of Aq-A-2 subhalos within 200 kpc (where all the stream particles examined are located) gives $\sigma=120$ km \unit{s}{-1}, predicting a width of 170 km \unit{s}{-1} for the interaction-speed distribution. We obtain slightly higher velocity dispersions than this for the interacting subhalos in our samples, but it is worth noting that the best-fit overall speed distribution is skewed substantially toward low velocities; if one instead matches the peak of the M-B distribution to that of the Aq-A-2 subhalos one obtains $\sigma=165$ km \unit{s}{-1} for the subhalo velocity dispersion and hence a predicted width of about 230 km \unit{s}{-1} for the interaction speed distribution, which is more in line with what we find. It is also worth noting that the recorded distributions of encounter speeds are biased towards low-speed encounters since these remain in the encounter volume longer, and thus are more likely to be recorded in a snapshot. For a fixed threshold distance this bias is roughly proportional to $1/\sub{v}{rel}$, and is also direction-dependent---subhalos moving parallel to the stream are more likely to be detected than those moving perpendicularly. We will discuss this bias, related to the time-resolution of the simulation, in more depth in Section \ref{subsec:resolution}. On the other hand, stream stars can reach speeds of up to $\sim 300$ km \unit{s}{-1} near pericenter (as seen in the right-hand panel of Figure \ref{fig:OrphanXYsnaps}), and the subhalo number density increases toward the center of the host halo, which may offset the time-resolution bias somewhat. Most importantly, the broad width and high mean speed of the subhalo distribution results in tails at both very high and very low encounter speeds. In general the lower the relative speed, the more disruptive the interaction will be; Figure \ref{fig:speedDist} confirms that the low-speed tail extends down to very low relative speeds (under 100 km \unit{s}{-1}) for even the lowest-mass subhalos considered.

\subsection{Frequency and mass spectrum of interactions}
\label{subsec:freqmass}
Figure \ref{fig:enc-with-time-all} shows the aggregate time-evolution of the encounter rate for all the streams in the sample. In order to stack the stream encounter histories, we time-shift each individual stream's encounters relative to its formation time, and then divide by its age, before adding all the encounters together and rebinning. We can then compare the aggregate encounter rate to the predicted $t^2$ behavior (dotted gray lines). In the case of the 5 kpc threshold, there appears to be some base rate of random encounters (roughly one per time-bin on average) before $\sub{t}{form}$ so we start the power-law from a constant instead of zero at $t=\sub{t}{form}$. As observed for stream O in Figure \ref{fig:OrphanIntxAll}, at late times the interaction rate appears to fall below this power-law behavior for the fixed-distance thresholds, which we attribute to a drop in particle resolution. 

The mass-dependent thresholds have a much noisier signal, with the $r_{1/2}$ encounter rate consistent with noise and the $2r_{1/2}$ rate barely emerging from the noise level. Again, we attribute this to the low particle resolution of the tagged streams; the mass-dependent thresholds are generally well below 1 kpc so they are the most susceptible. For this reason, we decided to only consider aggregate results from the $2r_{1/2}$ threshold and not quote any results from the $r_{1/2}$ threshold since it is so undersampled.

\begin{figure*}
\includegraphics[width=\textwidth]{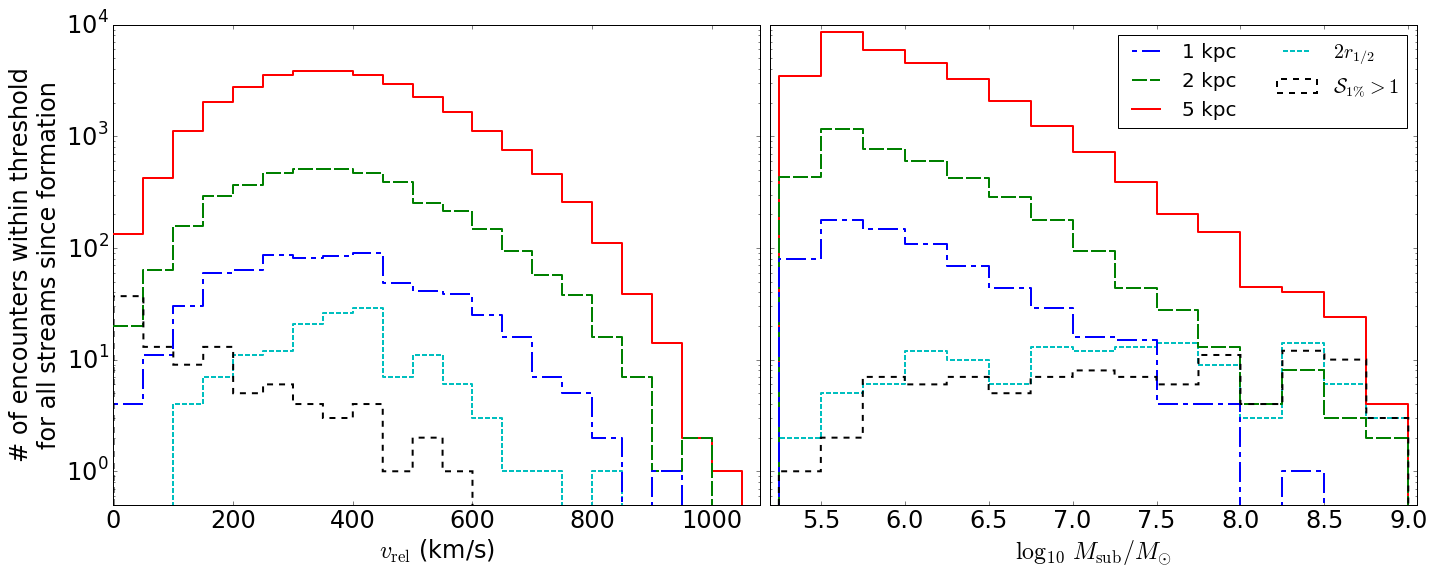} 
\caption{Distributions of relative velocity (left) and total mass (right) for all the encounters with the thin streams, for the three distance thresholds and the mass-dependent $2r_{1/2}$ threshold, and for all encounters within 5 kpc with $\mathcal{S}_{1\%}>1$. }
\label{fig:mass-and-vrel}
\end{figure*}

\begin{figure*}
\includegraphics[width=\textwidth]{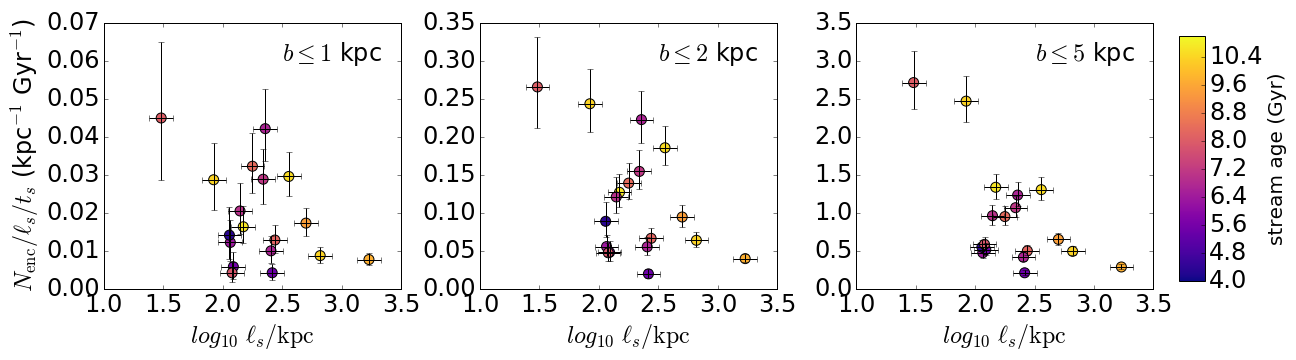} 
\caption{Rate of subhalo encounters (Equation \eqref{eq:rate}) within the three fixed distance thresholds for the 18 streams in our sample. Horizontal error bars are 10\% of the estimated stream length; vertical error bars include the length uncertainty and Poisson uncertainty on the number of encounters, added in quadrature.}
\label{fig:encounterRate}
\end{figure*}

Figure \ref{fig:mass-and-vrel} shows the distributions of total mass and relative speed for all the encounters as a function of threshold. We also include the distributions of all encounters with $\mathcal{S}_{1\%}>1$ for comparison. The relative velocity distribution is similar for the fixed-distance and mass-dependent thresholds, while the strong velocity-dependence in the definition of $\mathcal{S}_{1\%}$ is apparent here. As expected, the three fixed-distance thresholds sample the overall mass distribution while the mass-dependent threshold shows the competition between the subhalo mass function and the scaling of subhalo size with mass. The distribution of encounters with $\mathcal{S}_{1\%}>1$ is similar to the mass-dependent threshold, which justifies our use of the half-mass radius to represent the ``sphere of influence" of a subhalo. We can conclude from these results that while the typical halo interacting with a stream will have a low mass, the masses of subhalos that come close enough to significantly affect the stream are fairly uniformly distributed over several decades in mass: a $10^6\ M_\odot$ subhalo is roughly as likely to have a strong interaction with a stream as a $10^8\ M_\odot$ subhalo.

Finally, using the measured lengths $\ell_s$ and ages $t_s$ of the streams, determined as described in Section \ref{subsec:lengthAge}, we computed the length-normalized interaction rate,
\begin{equation}
\label{eq:rate}
\eta \equiv \frac{\sub{N}{enc}}{\ell_s t_s},
\end{equation}
for each stream in the sample and for each of the constant distance thresholds. As discussed in Section \ref{subsec:freqmass}, we do not computer individual rates for the mass-dependent thresholds since the frequency of interactions is consistent with noise. Figure \ref{fig:encounterRate} shows the results. Although the streams vary widely in length, rates for the large group of streams with lengths of a few hundred kpc do not show length- or age-dependence in their encounter rate (though there is a large scatter). The large range of stream lengths (nearly a factor of 100) can apparently account for much of the equally wide range in the number of encounters shown in Figure \ref{fig:encounterRate}. Even after normalization, the highest encounter rates tend to be for the shortest streams and the lowest encounter rates for the longest streams. This effect is most likely an indication of our limited numerical resolution and our approximate method for measuring stream lengths, which can be greatly affected by the inclusion/exclusion of single particles in the limit of low resolution.  As an example of how these effects can combine, the streams with the two highest rates in the 5 and 2 kpc panels are the two shortest streams in the sample: both are near apocenter at present day (i.e. at the shortest length of any orbital phase). However, both streams also have a fairly large number of particles in a diffuse region, surrounding a more concentrated nucleus, that were excluded as outliers during the length determination. The phase-dependence of the length and the uncertainty in the length measurement itself, which in this case is probably skewed to the short side, combine to produce very high rates for these two streams.

Taking the median of our sample of 18 streams, we find that a stream that is 10 Gyr old and 10 kpc long at present day should have experienced $61.8^{+211}_{-40.6}$ subhalo encounters within 5 kpc, $9.1^{+17.5}_{-7.1}$  encounters within 2 kpc, and $1.5^{+3.0}_{-1.1}$ encounters within 1 kpc. The error bounds given here are the extrema of the range of encounter rates over the streams we tracked.

\begin{figure}
\includegraphics[width=0.45\textwidth]{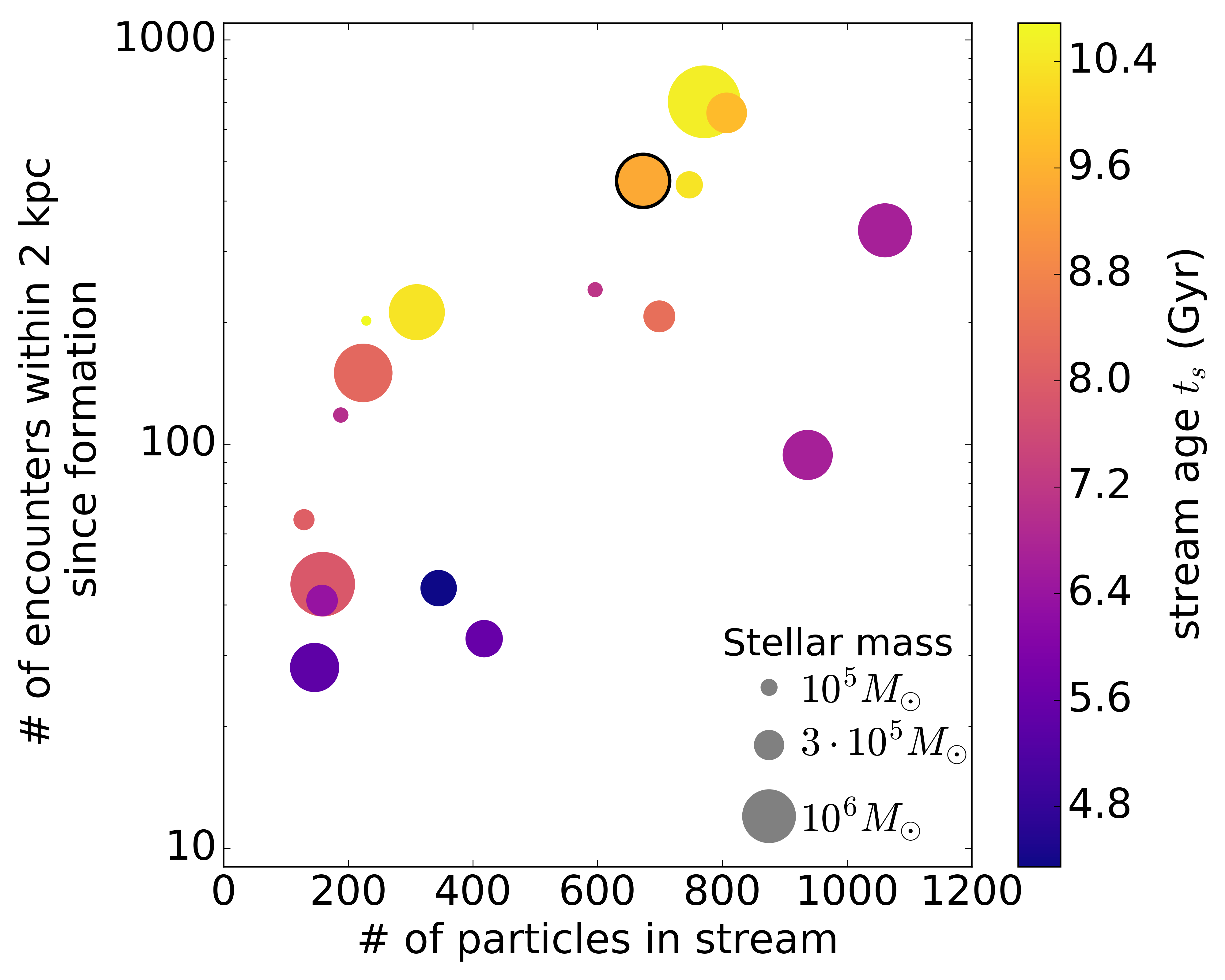} 
\caption{Number of encounters within 2 kpc versus number of particles in stream. Size and color denote the stellar mass of the stream progenitor and age of the stream (defined as in Section \ref{subsec:lengthAge}), respectively. The point circled in black is stream O. }
\label{fig:resolution-effects}
\end{figure}

\subsection{Resolution effects}
\label{subsec:resolution}

Encounters with the stream are undersampled because of the low time-resolution of the snapshots: for $t\gtrsim 1$ Gyr, the time between snapshots is 0.1546 Gyr, so a subhalo with a typical velocity of ~130 km/s moves about 20 kpc between snapshots. The undersampling is less severe for larger threshold distances, and is not strictly proportional to the single-particle threshold distance since we look for subhalos within that distance of \emph{any} particle in the stream. As the stream gets longer, this effect is mitigated somewhat as the stream length approaches the average distance traveled by a subhalo between two snapshots. Thus the slope at which the interaction rate increases at middle times in Figure \ref{fig:OrphanIntxAll} may be slightly steeper than if we had infinitely good time-resolution. 

The decrease in the slope of the interaction rate toward the end of the simulation (evident for Stream O in Figure \ref{fig:OrphanIntxAll} and still marginally present in the aggregated history of Figure \ref{fig:enc-with-time-all}) is probably an effect of particle resolution. Toward the end of the simulation when the tagged stream particles are very spread out, a lack of resolution can cause us to miss subhalos that would have interacted with the stream. Figure \ref{fig:resolution-effects} illustrates this point for our entire sample: the number of encounters is primarily correlated with the number of particles in the stream, whether one considers a fixed threshold (left) or a mass-dependent one (right). 

Time- and particle resolution both affect our results in the same way, so the results we present in this work should therefore underestimate the number of interactions between a stream and dark substructures. We think that particle resolution is probably the more seriously limiting of the two issues based on the scaling of the median interaction rates for the 18 streams. If we had perfect time-resolution, the median rates for the fixed-distance threshold should scale as the cube of the threshold length: $1:8:125$ for the 1:2:5 kpc thresholds. Our measured rates scale as $1:6.1^{+11.6}_{-4.8}:41.2^{+141}_{-27.1}$, so the medians are a bit on the low side but the proper scaling is within the range we obtain for different streams. 

The mass resolution of the simulation mainly affects our ability to detect repeat encounters, which would likely be interactions with subhalos that accompanied the progenitor of the stream during infall. The dark-matter halos of the stream progenitors are not very massive, however, and if a stream progenitor is the largest halo in an infall group, then its most massive companion is likely to be only about 1/100th as massive, and hence will be resolved with about as many particles as are tagged for the stellar stream. For a progenitor like that of our example O stream, the most massive companion will only be resolved with about 700 particles. It takes about 100 particles to satisfactorily resolve a subhalo, so only the few most massive companion subhalos to a stream progenitor, that might give rise to multiple encounters, will be resolved in the simulation. Thus the frequency of repeat encounters is likely also underestimated by this work.

\section{Discussion}
\label{sec:discussion}

\begin{figure}
\includegraphics[width=0.45\textwidth]{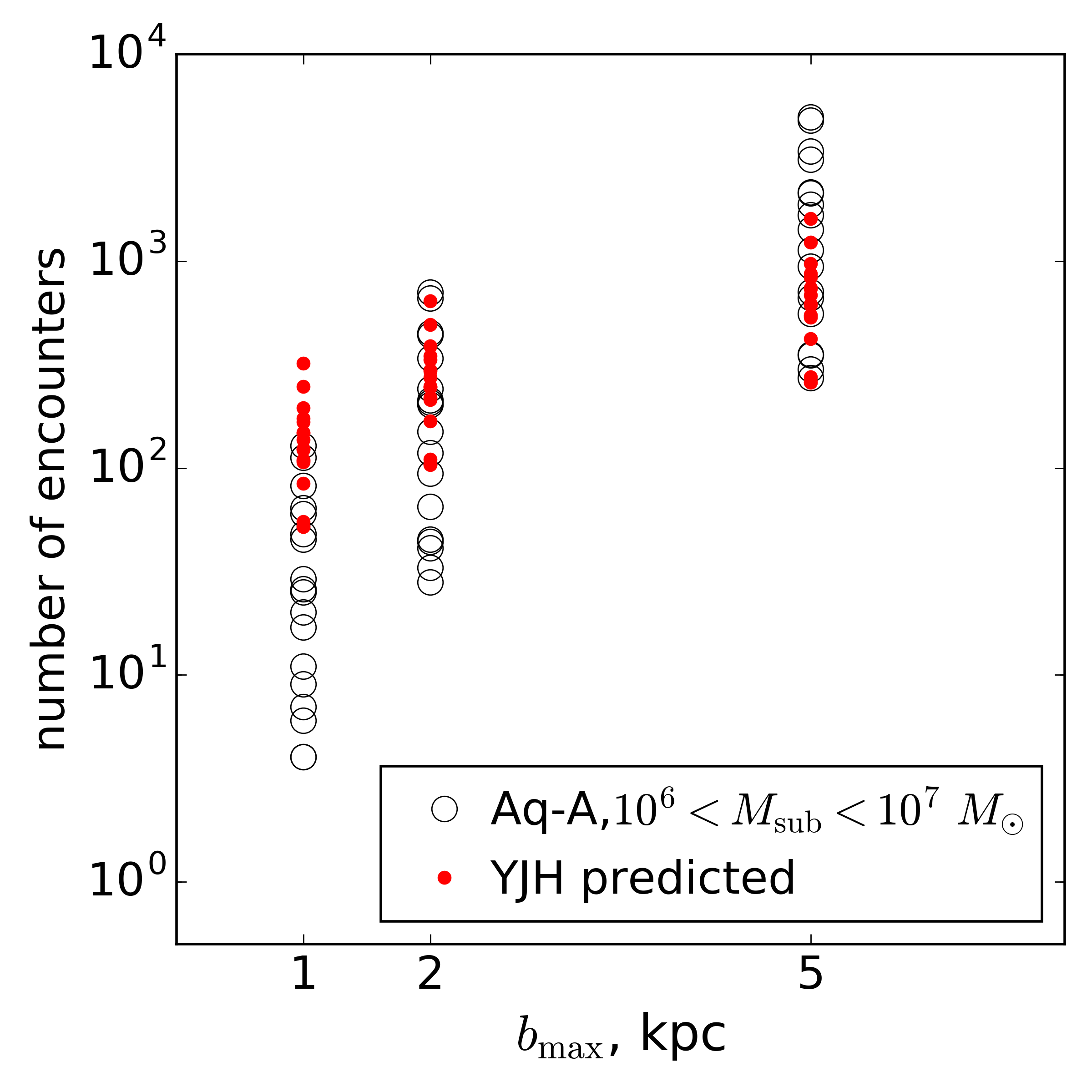}
\caption{Comparison between our measured numbers of encounters within different distance thresholds (black circles) and analytic estimates based on \citet[][Equation \ref{eq:yjh}]{Yoon2011} (red points). See text for the details of how the predictions were calculated.}
\label{fig:comparison}
\end{figure}

The encounter rate that we measure can be compared to analytic
expressions in the literature. Figure \ref{fig:comparison} compares our results to the estimate derived by \citet{Yoon2011} for the total number of encounters in the mass range $10^6-10^7\ M_{\odot}$ within a constant impact parameter $\sub{b}{max}$, based on the subhalo density in the Via Lactea II simulation (their Equation 16):
{\small 
\begin{eqnarray}
\sub{N}{enc}(10^6<\sub{M}{sub}<10^7\ M_{\odot}) =  20 \left(\frac{\sub{R}{circ}}{13.7 \textrm{ kpc}}\right)\left(\frac{\sub{b}{max}}{0.58\ \textrm{kpc}}\right) \qquad  \nonumber   \\
 \times \left(\frac{\sigma}{120 \textrm{ km}\unit{s}{-1}} \right) \left(\frac{t}{8.44\ \textrm{Gyr}}\right)^2 \left(\frac{\sub{n}{sub}}{0.0006\ \unit{kpc}{-3}}\right) \left(\frac{\Delta\Psi}{1^\circ}\right)\left(\frac{0.55\ \textrm{Gyr}}{T_{\Psi}}\right). 
\label{eq:yjh}
\end{eqnarray}
}
Given that the VLII halo is quite similar to Aq-A-2 in mass, we use the fiducial values for the stream lengthening rate per orbit $\Delta\Psi$, and the orbital period $T_\Psi$. We also take the fiducial value of the subhalo density $\sub{n}{sub}$; although this value is calculated at the low end of our streams' distance range (13.7 kpc), the Aq-A-2 halo has slightly more substructure than VLII so the subhalo density at 10-60 kpc is actually about the same as VLII at 13.7 kpc \citep{Springel2008}. We use a subhalo velocity dispersion $\sigma=139.3$ km \unit{s}{-1}, which is the best-fit value from a fit to a Maxwell distribution: as we discuss in Section \ref{subsec:vrel}, this functional form is not a very good fit to the real subhalo velocity distribution, but the model of YJH presumes this form so we use it for consistency. The predictions depend only linearly on the velocity dispersion, and more strongly on other things like the age of each stream, which varies among the different streams, whereas $\sigma$ is the same for all of them and so will only shift the predictions up or down by some constant. Additionally the best-fit value does approximate the width of the distribution (perhaps slightly over-estimated thanks to skewness that cannot be accommodated by the model). For these reasons we do not expect the poor fit to have a strong effect on the result.  Plugging in each stream's median radius for $\sub{R}{circ}$, each stream's age $t_s$ as $t$, and using the different constant distance thresholds as \sub{b}{max} produces the predictions shown as red points, which we compare to the measured number of encounters within the same thresholds (black circles). At small \sub{b}{max} our rates are lower by a factor of about 10, while at large \sub{b}{max} the predictions agree; this is partially expected behavior given the resolution limits since wider thresholds are less sensitive both to the time between snapshots and increasing distance between particles. On the other hand, the orbital period and lengthening rate for our streams is likely longer than the fiducial value since many are on orbits at larger distances in the halo (VLII is also a bit more massive than Aq-A-2), which may also contribute to the larger rates calculated by \citeauthor{Yoon2011}. 

\begin{figure}
\begin{center}
\includegraphics[width=0.45\textwidth]{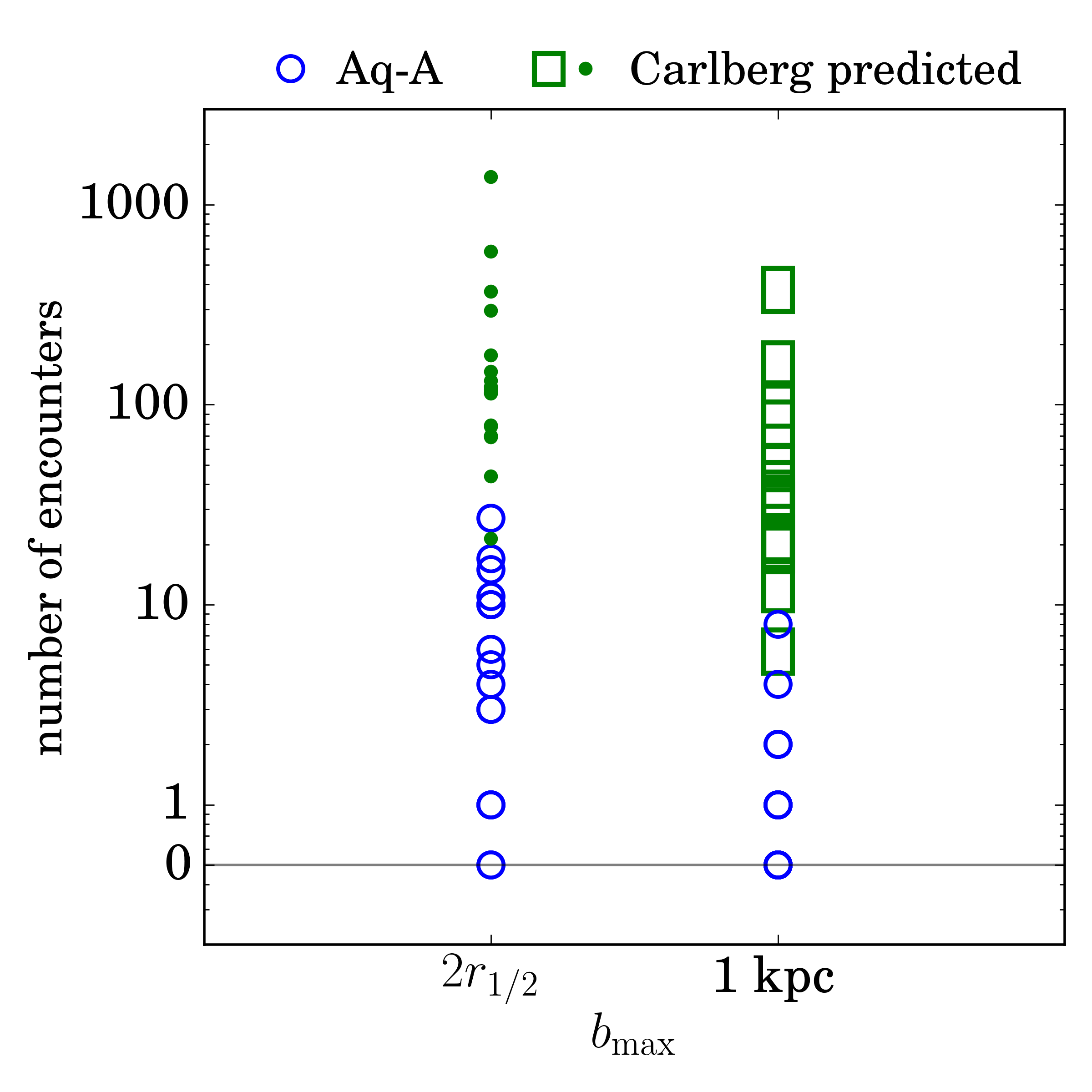}
\caption{Comparison between our measured numbers of encounters within different distance thresholds and mass ranges (blue circles) and analytic predictions based on \citet{Carlberg2012}. In the left-hand column, all interactions within $2r_{1/2}$ with $M>10^6M_\odot$ are compared with the predicted rate of Equation \eqref{eq:carlberg} presuming that any interaction within this distance will open a discernible gap. In the right-hand column, interactions with $b<1$ kpc and $M>10^{7.28}M_\odot$ are compared with the rate predicted by Equation \eqref{eq:carlberg}, based on applying Carlberg's empirical criterion for opening a gap as described in the text.} 
\label{fig:comparison_ray}
\end{center}
\end{figure}

\citet{Carlberg2012} also derives an expression for the rate of subhalo interactions with a slightly different approach. Using constrained simulations they determine the maximum impact parameter required to open a gap in a stream of a fixed density contrast relative to the unperturbed stream, as a function of the time since the interaction, the mass of the perturbing subhalo, and the orbital radius $r$. Because our mass-dependent thresholds are heavily subject to undercounting, we sought instead to try to identify the set of subhalos within a fixed encounter threshold that satisfy this criterion, which we computed for each stream in our sample using its mean orbital distance for $r$ and its age $t_s$ for the time since interaction. This is an upper limit on the required encounter distance, since interactions that happened later on would need to come closer to produce a gap of the same depth in a shorter time. We find that based on this reasoning, our 1 kpc fixed threshold corresponds to the required encounter distance for subhalos in the mass range $10^{7.28} - 10^{8}\ M_{\odot}$. There are 22 encounters within 1 kpc within this mass range, over a total of 18 streams; the number of qualifying encounters per stream ranges from zero to 8. 

To compare with Carlberg's prediction we first tried simply applying Equation 5 of \citet{Carlberg2012},
{\small
\begin{equation}
\frac{\mathcal{R}_{\cup}(\hat{M}, r)}{\unit{kpc}{-1} \unit{Gyr}{-1}} = 0.0066\left(\frac{r}{100 \textrm{ kpc}}\right)^{0.23}\left( \frac{n(r)/n_0}{6} \right)  \left(\frac{\sigma}{120 \textrm{ km}\unit{s}{-1}}\right) \left(\frac{\hat{M}}{10^8\ M_{\odot}}\right)^{-0.35},
\label{eq:carlberg}
\end{equation}
}
which uses fiducial values taken from the subhalo distribution in the Aquarius
halos. We used the median distance of star particles in each stream as
the galactocentric distance $r$ and set the subhalo velocity dispersion $\sigma$ to 139.3 km \unit{s}{-1} as with the YJH estimate. The scaled subhalo number density
$n(r)/n_0$ is obtained for Aq-A-2 from Figure 11 of
\cite{Springel2008}, and is about 15 for the range explored by our
sample streams (10-60 kpc). Based on our application of the gap-opening criterion, we calculated upper and lower limits on $\mathcal{R}_{\cup}$ for each stream by taking $\hat{M} = 10^{7.28}$ and $\hat{M} = 10^{8}$ as the range of possible minimum masses to open a gap given 1 kpc maximum impact parameter. We then multipiled Equation \eqref{eq:carlberg} by each
stream's measured length $\ell_s$ and age $t_s$ to get an estimate of the expected number of encounters in each stream, to compare with the 22 encounters in this same mass range over the 18 streams. This comparison is shown in the right-hand column of Figure \ref{fig:comparison_ray}, with the range of the predicted number of encounters denoted by the green boxes. This approach presumes that each stream stays at its present-day length over its entire formation time, whereas in reality they grow roughly linearly with time. Furthermore it presumes a circular orbit whereas the actual stream orbits are fairly eccentric and spend most of their time at larger distances where the subhalo density is lower. Added to the fact that for this distance threshold we are likely missing some events due to particle and time resolution, Equation \eqref{eq:carlberg} should produce an over-estimate relative to what we observe and we find that this is so; in fact the over-estimate is something like a factor of 10 to 50. However, the number of scattering events we detect in this mass range is very small, so it is difficult to quantify by how much this approach over-estimates the number of encounters.

One could also construe Equation \eqref{eq:carlberg} in terms of the mass-dependent criteria we use to count events within $r_{1/2}$ or $2r_{1/2}$, since this criterion is another way to select interactions that approach close enough to open a gap. Carlberg finds from his simulations that for a discernible gap to open from an interaction 7 Gyr in the past (the fiducial age assumed for all his predictions) the perturbing subhalo must approach within 1-3 scale radii presuming an NFW profile. Satellite galaxies (and simulated subhalos) are tidally truncated beyond a few scale radii; for subhalos orbiting at a few times the scale radius of the parent halo (where our streams are located) this truncation radius is also about 1-3 times the scale radius \citep{Hayashi2003} so it is roughly consistent to use $2r_{1/2}$ as the gap opening criterion for the approach distance. We compare all encounters within $2r_{1/2}$ to the prediction of Equation \eqref{eq:carlberg} for a minimum mass of $\hat{M}=10^6\ M_{\odot}$: in this case all the events we record are gap-opening, so the minimum mass to open a gap is equal to the lowest mass in our sample, which we cut off at $10^6\ M_{\odot}$. This comparison is shown in the left-hand column of Figure \ref{fig:comparison_ray}. This set of interactions, although still undersampled for the reasons described above, has slightly more events in it than the mass-limited, 1 kpc version in the right-hand column, but the rates we measure are still much lower than predicted by Carlberg's formula. 

The typical stream in our sample has had only a handful of close interactions with subhalos in its lifetime (less than 10 within 2 kpc, and a 20\% chance of one within $2r_{1/2}$). Given this fairly low rate of encounters, it is at first glance surprising how disturbed the morphology of a stream can be: our example stream starts out as a completely smooth distribution but has some significant lumps and holes in it (both in position and energy) by the end of the simulation, yet we count only one interaction with a subhalo within $r_{1/2}$ over its entire lifetime. \citet{2015ApJ...803...75N} found similarly disorganized-looking streams in their simulations when starting from a smooth globular-cluster-like phase space distribution and using the Via Lactea simulation. In our case, the interaction rate is undercounted for a number of reasons, most notably the fairly large time between snapshots and the low resolution of the streams themselves, but also because we did select streams that still appear fairly coherent at present day, which automatically picks out ones that have had fewer disruptive encounters. The stream lengths on the other hand are probably over-estimated compared to what one would be able to identify as part of the stream observationally, since the ends of our simulated streams are often very diffuse. Additionally, gaps in streams will grow faster in a gravitational potential like Aq-A-2's, which is triaxial, than in a spherically-symmetric potential \citep{1999Helmi}, especially for streams on more eccentric orbits \citep{2015ApJ...808...15C}. Streams in less symmetric potentials are also intrinsically more complex-looking, even without the addition of substructure \citep[e.g.][]{2015ApJ...799...28P}, since these admit a wider range of orbit families than streams in spherical potentials. Gap size also varies significantly with orbital phase \citep{2016arXiv160608782H}, leading to a wider variety of effects. Given the range of possible gap growth rates, the variety of orbits and phases our streams explore, and the fact that we are missing some encounters due to time-resolution, it is perhaps not as surprising that the streams look more ``messed-up,'' and the gaps in them harder to identify and disentangle, than one might expect from our measured encounter rate.

\section{Summary and conclusions}
\label{sec:concl}

In this work we used a new, simplified tagging scheme to produce an accreted stellar halo from the Aquarius A-2 dark-matter-only simulation, and studied interactions between dark subhalos and thin stellar streams orbiting the main halo. We recorded encounters where a subhalo (identified by its most-bound particle) came within several threshold distances of any tagged star particle in a stream. We used three constant distance thresholds (1, 2, and 5 kpc) and two mass-dependent thresholds based on the half-mass radius of each subhalo ($\sub{r}{1/2}$ and $2\sub{r}{1/2}$).

First we tracked a single stellar stream, observing the rate at which
interactions occurred at different thresholds, the evolution of the
physical and phase-space shape of the stream over time, and the mass
and velocity spectra of the encounters. The initial phase-space
distribution of the tagged particles in the stream's progenitor is
smooth at the time of infall, yet we see significant structure,
including discontinuities in position and energy, arise as the stream
evolves. As the stream gets longer the number of encounters increases
and by late times, multiple subhalos are passing close to the stream
in every snapshot, and up to 26 within 2 kpc at a given point in time.  Only some of these passes bring the subhalo closer than its own half-mass radius: in the case of our example stream, only 12 subhalos pass within $2r_{1/2}$ and 2
within $r_{1/2}$ over its lifetime. Additionally, we find only one
case out of many where a subhalo experienced multiple encounters with
the stream, and the velocities of the interacting subhalos are not
significantly correlated with that of the stream.

Then we identified a sample of 18 stellar streams that still appeared
streamlike at the present day and repeated the tracking of encounters
for all streams in the sample. The growth of the number of encounters
with time over the entire sample confirms what we observed for the
single stream, although at late times low resolution likely causes us
to under-count interactions. The relative speeds of the interacting
subhalos are representative of the general subhalo population, as are
their masses when a fixed threshold distance is used. For the
mass-dependent threshold distances, subhalos in the range $10^6$ to
$10^8\ M_{\odot}$ (and perhaps up to $10^{8.5}$) are equally likely to have an interaction
with a stream within $2r_{1/2}$. 

Comparing the rate of interactions we measure to analytic predictions, we find that our measured rate is lower (as one would expect given the resolution limitations) and closer to these predictions for larger maximum impact parameter (as one would also expect for a resolution-limited rate). Our median measured rates for encounters, per 10 Gyr per 10 kpc, are: 62 within 5 kpc, 9 within 2 kpc, 1.5 within 1 kpc, and 0.2 within $2r_{1/2}$.

\begin{acknowledgements}
RES is supported by an NSF Astronomy and Astrophysics Postdoctoral Fellowship under award AST-1400989. RES and AH gratefully acknowledge support from the European Research Council under ERC-Starting Grant GALACTICA-240271. AH is also supported by a VICI grant from NWO.
\end{acknowledgements}

\bibliographystyle{aa}
\bibliography{ClumpyHalos}

\end{document}